\documentclass[11pt,en,cite=authoryear]{elegantpaper}
\usepackage{amsmath}
\usepackage{adjustbox}
\usepackage{lscape}
\usepackage{pdflscape} 
\usepackage{color,soul}
\usepackage{xcolor}
\usepackage{graphicx}
\usepackage{longtable}
\usepackage{verbatim}
\usepackage{array}
\newcolumntype{R}[1]{>{\raggedleft\arraybackslash}p{#1}}
\usepackage{bbm}
\usepackage{multirow}
\usepackage{microtype}
\usepackage{setspace}

\usepackage{comment}

    \title{Dissecting AI Trading: Behavioral Finance and Market Bubbles\thanks{Shumiao Ouyang, Sa\"{i}d Business School, University of Oxford, email: \email{shumiao.ouyang@sbs.ox.ac.uk}. \\Pengfei Sui, School of Management and Economics, The Chinese University of Hong Kong, Shenzhen, email: \email{psui@cuhk.edu.cn}. \\We thank Shang Wang and Dong Er for excellent research assistance. Shumiao Ouyang thanks Oxford RAST for their research support and Oxford Sa\"{i}d for funding support. All errors are our own.}}

\author{Shumiao Ouyang \and Pengfei Sui}

\usepackage{array}

\begin{document}
\doublespacing

\maketitle

\begin{abstract}
We study how AI agents form expectations and trade in experimental asset markets. Using a simulated open-call auction populated by autonomous Large Language Model (LLM) agents, we document three main findings. First, AI agents exhibit classic behavioral patterns: a pronounced disposition effect and recency-weighted extrapolative beliefs. Second, these individual-level patterns aggregate into equilibrium dynamics that replicate classic experimental findings (\cite{smith1988bubbles}), including the predictive power of excess demand for future prices and the positive relationship between disagreement and trading volume. Third, by analyzing the agents' reasoning text through a twenty-mechanism scoring framework, we show that targeted prompt interventions causally amplify or suppress specific behavioral mechanisms, significantly altering the magnitude of market bubbles.
\end{abstract}

\bigskip 


\clearpage
\clearpage

\section{Introduction}





Artificial intelligence is rapidly transforming financial markets, with Large Language Models (LLMs) increasingly deployed in asset pricing, portfolio allocation, and algorithmic trading. As these AI agents assume a greater role in price discovery, understanding their behavioral patterns, how these patterns aggregate into market-level dynamics, and whether they can be controlled has become a first-order question for financial stability.

We address this challenge by designing a simulated asset market populated entirely by autonomous AI traders, allowing us to study their behavior within the framework of economic equilibrium. By adopting the classic open-call auction paradigm of \cite{smith1988bubbles}, we establish a rigorous, controlled laboratory for observing endogenous price formation. This design closely mimics the institutional features of real-world exchanges, enabling us to trace the causal chain from micro-level trading behaviors directly to aggregate market dynamics.

Leveraging this simulated market, we analyze both the micro-level properties of AI traders and the resulting macro-level equilibrium dynamics. A natural question is why AI traders would exhibit human-like behaviors in a purely digital environment. We hypothesize that because LLMs are trained on vast, human-generated datasets—including historical financial narratives, investor sentiment, and documented market reactions—they likely internalize human cognitive patterns. Viewed through this lens, AI trading decisions serve as highly sophisticated simulations drawn from a comprehensive repository of human behavior.

Our empirical analysis documents findings that are, in many ways, strikingly consistent with classical behavioral finance studies on human investors. At the micro-level, we find that AI agents exhibit pronounced behavioral tendencies. First, they demonstrate a strong disposition effect, systematically selling assets that have gained in value while holding onto losing positions. Second, they form highly extrapolative beliefs, forecasting future prices based disproportionately on recent past returns.

We further examine the connection between these stated beliefs and actual trading behaviors. While previous literature on human retail investors (e.g., \cite{giglio2021}) documents a surprisingly weak link between expectations and portfolio allocations—often attributed to real-world frictions like psychological inertia, transaction costs, or inattention—we find that AI traders operate differently. Unburdened by these human frictions, AI agents exhibit a remarkably tight coupling between their stated expectations and their executed trades.

Turning to equilibrium dynamics, we demonstrate how these individual AI heuristics aggregate into familiar macro-level market phenomena. First, we document a strong correlation between the cross-sectional disagreement in AI beliefs and aggregate trading volume, directly mirroring the theoretical mechanisms of heterogeneous beliefs in human markets. Second, replicating the seminal findings of \cite{smith1988bubbles}, we show that the bid-offer gap—acting as a measure of excess demand—serves as a robust leading indicator for future price changes, driving the adaptive trajectory of market prices.

To move beyond correlative outcomes, we leverage the unique observability advantage of LLMs to provide direct evidence for the role of behavioral theories. By extracting the textual reasoning generated by the AI agents, we first validate that these self-revealed narratives are systematically consistent with their quantitative beliefs and trading actions. We then analyze the differences in reasoning during bubble versus non-bubble episodes. We find that during bubbles, agents explicitly articulate speculative strategies, such as momentum chasing and riding the bubble.

Finally, we show that implications from behavioral finance can be broadly utilized to actively recalibrate AI agents and stabilize market equilibrium.  By designing targeted prompts that suppress specific behavioral biases, we significantly reduce the magnitude of market bubbles. To verify that the direction of the intervention drives this outcome, we introduce contrasting prompts designed to explicitly amplify these patterns; as expected, this reverse intervention exacerbates bubble formation. This programmability offers profound policy implications, suggesting that regulators could implement "cognitive guardrails" at the prompt level to foster systemic stability in AI-driven financial markets.

Our paper makes three primary contributions, bridging behavioral finance and the emerging study of artificial intelligence in economic systems. First, we open the black box of AI trading by documenting that LLM agents exhibit pronounced consistency with patterns established in the behavioral finance literature. Specifically, we show that AI agents display a robust disposition effect (\cite{shefrin1985disposition}, \cite{odean1998investors}, \cite{ouyang2025fixed}) and form extrapolative expectations (\cite{greenwood2014}, \cite{barberis2018extrapolation}, \cite{jin2022asset}). Furthermore, while the existing literature documents a weak empirical link between stated beliefs and portfolio allocations in human investors (\cite{giglio2021}), we document that AI agents exhibit a notably tighter belief-action coupling, where stated expectations translate more directly into trading behavior.

Second, we contribute to the broader behavioral finance literature by establishing a clear micro-to-macro transmission channel. Building on the canonical experimental asset market tradition of \cite{smith1988bubbles} (see \cite{palan2013review}), we demonstrate how individual-level heuristics aggregate into endogenous market dynamics. While the Smith, Suchanek, and Williams (SSW) design has long served as a laboratory for human bubbles, we adapt this open-call auction framework to trace the causal chain from micro-level AI cognition to macro-level price formation. By systematically auditing the agents' textual reasoning through a multi-mechanism scoring framework, we provide direct evidence of the cognitive mechanisms driving aggregate market outcomes.

Third, we advance the rapidly emerging literature on large language models as economic agents (\cite{horton2023large}, \cite{mo2025generative}). Previous research has documented isolated behavioral biases in LLMs, such as prospect-theoretic violations and extrapolation errors (\cite{ouyang2024ethical}, \cite{chen2024does}, \cite{ross2024llmeconomicusmapping}, \cite{bini2025behavioral}). Concurrently, several studies have begun simulating LLMs in asset markets, benchmarking them against humans (\cite{henning2025llm}), exploring market microstructure (\cite{lopezlira2026trade}), and analyzing participation in sequential bubble games (\cite{cartea2026aibubbles}). We differentiate our work from these concurrent efforts by moving beyond observation to causal intervention. By demonstrating that targeted prompt interventions can causally amplify or suppress specific behavioral mechanisms—thereby significantly altering bubble magnitudes—we offer a novel and actionable policy lever for the regulation of AI-driven financial markets.

The paper proceeds as follows. Section \ref{sec:data_methodology} details the experimental design, encompassing the economic environment, agent architecture, and equilibrium mechanism. Section \ref{sec:behaviors} documents individual trading behaviors—such as the disposition effect, extrapolative expectations, and belief-action coupling—while Section \ref{sec:equilibrium} analyzes the resulting equilibrium price dynamics. Section \ref{sec:reasoning} explores the cognitive mechanisms revealed through the agents' reasoning text. Section \ref{sec:exogenous_manipulation} demonstrates how targeted prompt interventions can causally amplify or suppress these behavioral mechanisms. Finally, Section \ref{sec:discussion} addresses interpretive limitations, and Section \ref{sec:conclusion} concludes.

\section{Experimental Design}
\label{sec:data_methodology}


To systematically investigate AI decision-making in financial environments, we design a simulated asset market adapting the classic experimental finance paradigm ( \cite{smith1988bubbles}; \cite{henning2025llm}). This controlled setting provides a rigorous framework grounded in economic equilibrium, offering a common language to address fundamental questions about AI agent behaviors, price formation and market efficiency.\footnote{We use ``AI agent'' in general discussion and ``LLM agent'' when emphasizing the specific large language model architecture. Both terms refer to the same autonomous trading entities in our experiments.} Crucially, it allows us to simultaneously observe micro-level activities—such as internal reasoning, belief formation, and trading decisions—and macro-level dynamics, including equilibrium prices, trading volumes, and endogenous bubbles or crashes. Consequently, we generate a rich, multi-layered panel dataset linking cognitive processes directly to market outcomes.

\subsection{Economic Environment}\label{subsec:market_setup}

Drawing upon the experimental asset market framework pioneered by \cite{smith1988bubbles}, the simulated market operates as a multi-period open-call market. The trading horizon consists of 3 initial practice periods followed by \(T = 20\) main trading periods. The economic environment features two available assets: a risk-free asset (cash) and a risky asset (stock). 

At the onset of the experiment, each AI agent is endowed with an identical initial portfolio consisting of 100 units of cash and 4 shares of the risky asset. Inventories of both cash and shares are endogenous and carry over from one period to the next. These assets yield returns according to the following structure:
\begin{itemize}
    \item \textbf{Risk-Free Rate:} Uninvested cash balances held at the end of each trading period accrue interest at a fixed rate of \(r = 5\%\).
    \item \textbf{Risky Dividend:} Each share of the risky asset pays a stochastic dividend at the end of each period. The dividend is drawn from a two-point discrete distribution, yielding either a low dividend of 0.4 units or a high dividend of 1.0 unit with equal probability. Consequently, the expected per-period dividend is \(\mathbb{E}[D] = 0.7\) units. The probabilistic nature of this dividend structure is common knowledge among all agents.
\end{itemize}

In standard finite-horizon asset experiments, the intrinsic value of a share systematically declines as the number of remaining dividend-paying periods decreases. To eliminate this declining trajectory and induce a strictly constant fundamental value, we implement a terminal buyout condition (analogous to the buyout treatments in \cite{smith1988bubbles}). At the conclusion of the 20th trading period, all outstanding shares of the risky asset are automatically redeemed for a fixed terminal value. 

Because of this terminal buyout and the presence of a risk-free alternative, the fundamental value (\(FV\)) behaves as the present value of a perpetuity. It is determined by the rationally expected future dividends relative to the risk-free interest rate:
\begin{equation}
    FV = \frac{\mathbb{E}[D]}{r} = \frac{0.7}{0.05} = 14
\end{equation}

This constant fundamental value of 14 units serves as an objective, time-invariant rational expectations benchmark. By establishing a clear no-arbitrage price, this design enables us to cleanly classify macro-level market phenomena---such as price bubbles and crashes---and rigorously evaluate the rationality of the agents' endogenous trading behavior.

\subsection{AI Agent Architecture}
\label{subsec:agent_architecture}

The core component of our experimental framework is the autonomous AI agent. To emulate realistic market participation, we provide detailed prompt instructions that inform the agents about the market environment and prevailing trading rules. Throughout all sessions, agents are explicitly instructed to maximize their individual earnings while remaining strictly unaware of their counterparties' identities. For the complete set of prompts and instructions, please refer to Appendix \ref{sec:appendix_instructions}. 

In real-world applications, AI agents are powered by a variety of Large Language Models (LLMs). Due to underlying algorithmic and architectural differences, these LLMs exhibit distinct reasoning processes and behavioral patterns. To ensure our findings broadly reflect the current landscape of AI-driven trading, we select fourteen frontier large language models (see Table \ref{Table_bubble_stat} for the full list) to serve as the cognitive engines for our agents. 

Using these selected models, we first establish baseline environments, which we dub \textit{Single-Model Markets}. In these sessions, the market is populated by 20 agent instances, all driven by the exact same LLM. This controlled setting allows us to isolate the trading behaviors, cognitive biases, and resulting market equilibria intrinsic to each specific model.

To capture more complex market dynamics, we also implement \textit{Mixed-Model Markets}. In these environments, the market is populated by 24 agents evenly split between two different LLMs. For instance, in a mixed market pairing DeepSeek and GPT, 12 agents are powered by DeepSeek and the remaining 12 by GPT. Incorporating these mixed markets not only broadens the external validity of our study but also allows us to investigate behavioral heterogeneity and strategic differences across AI types. Furthermore, it enables us to analyze the interactions between different classes of investors—a setup that conceptually mirrors classical finance frameworks, such as the dichotomy between rational investors and noise traders.

The landscape of the market configurations is detailed in Table \ref{Table_bubble_stat}.

\subsection{The Decision-Making Process of AI Agent}
\label{subsec:agent_decisions}

\begin{figure}[p]
\centering
\caption{Simulation Process Within Each Trading Round}\label{Figure_process}
\includegraphics[width=\textwidth]{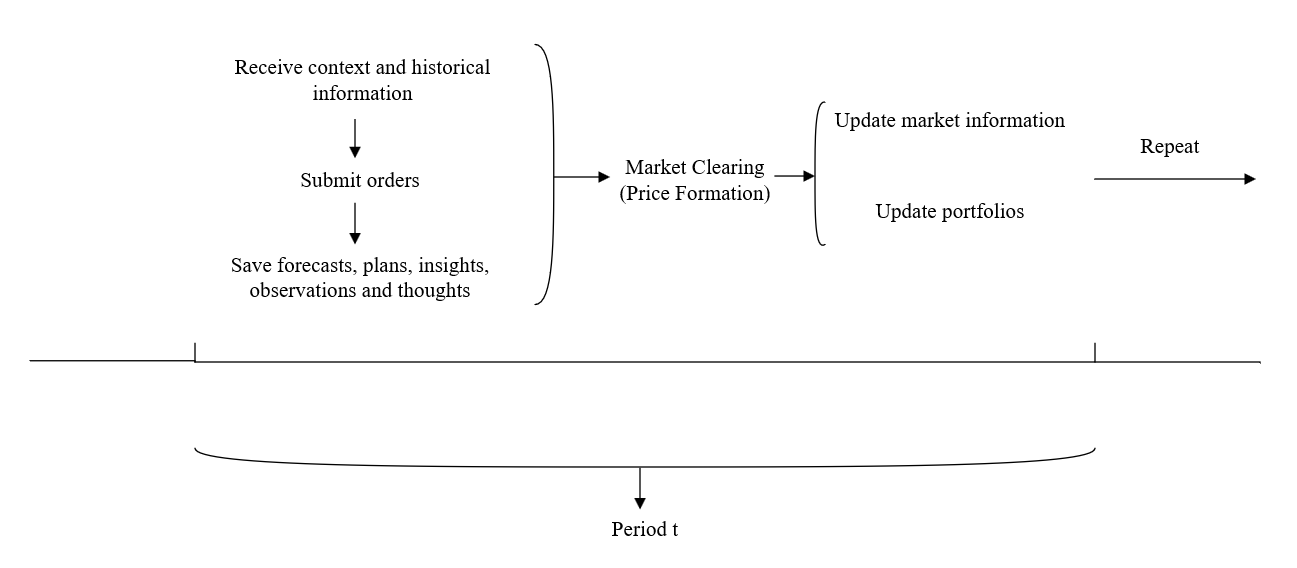}
\vspace{4pt}

{\footnotesize \noindent \textit{Note:} This figure illustrates the sequence of operations within a single trading round. At the beginning of each round, agents receive the current market state together with historical information. Based on this context, each agent submits orders and produces a structured response containing price forecasts, plans, insights, and accompanying reasoning. Submitted orders are aggregated through a market-clearing mechanism to determine the transaction price. The resulting price and executed trades are then used to update both the market state and individual portfolios. The updated state is fed into the next round, forming a closed-loop process that repeats over time.}
\end{figure}

Figure \ref{Figure_process} illustrates the sequence of operations within a single trading round. At the beginning of each round, agents receive the current market state together with historical information. Based on this context, each agent produces a structured response containing strategic reasoning and reflections, price forecasts for multiple horizons, and trading orders. The following subsections detail each component of this decision pipeline.

\subsubsection{Memory and Reasoning Extraction}
\label{subsec:memory_reasoning}

A fundamental methodological challenge in deploying LLMs within multi-round strategic environments is maintaining memory continuity and eliciting latent reasoning. To address this, we implement a structured Chain-of-Thought (CoT) architecture. At the onset of each round, agents receive a comprehensive state update formatted in JSON. This payload details historical market prices, aggregate trading volumes, the agent's own execution history, and their current portfolio balances. 

Importantly, prior to formulating any numerical trading decisions or price forecasts, agents are required to reveal their trading strategies and reasoning by updating two private text files:
\begin{itemize}
    \item \textbf{\texttt{PLANS.txt}:} A forward-looking document where agents articulate their intended trading strategies for upcoming rounds.
    \item \textbf{\texttt{INSIGHTS.txt}:} A reflective document where agents record observations regarding market dynamics, past errors, and learned heuristics.
\end{itemize}
These files are passed sequentially from round to round, serving as a persistent, evolving memory trace. Furthermore, following the initial practice rounds, agents complete a ``Practice Reflection'' to distill early lessons, effectively mimicking the learning phase observed in human-subject experiments. This architecture forces the models to explicitly articulate their internal cognitive state prior to action, yielding a rich textual dataset for subsequent behavioral analysis.

\subsubsection{Belief Elicitation: Price Forecasting}
\label{subsec:forecasting}

Following the reasoning phase, we implement an incentivized forecasting task to rigorously assess the rationality of agent expectations and test for extrapolative belief formation. Alongside their order submissions in period \(t\), agents are required to forecast the market-clearing price for the current period (\(t\)), as well as for future horizons \(t+2\), \(t+5\), and \(t+10\). 

Let \(f^{(t+h)}_{i,t}\) denote the forecast made by agent \(i\) in period \(t\) for the price at period \(t+h\), where \(h \in \{0, 2, 5, 10\}\). To ensure these forecasts reflect genuine economic beliefs rather than arbitrary outputs, agents are incentivized with a reward of 5 additional cash units at the end of the experiment for every forecast that falls within \(\pm 2.5\) units of the realized market price. Finally, to prevent formatting hallucinations and extreme outliers, forecasts are constrained to non-negative integers and bounded by dynamically updating upper limits (e.g., twice the current market price for near-term forecasts).

\subsubsection{Trading Orders}

In each trading round \(r\) of session \(s\), agents simultaneously submit limit orders to buy or sell shares. An order from agent \(i\) is defined as a price-quantity tuple, \((p, q)\). Agents may submit multiple orders per round. Following the standard constraints of experimental asset markets, trading is strictly limited by the agents' current cash endowments (working capital) and share inventories; margin purchasing and short selling are strictly prohibited.

\subsection{Equilibrium}
\label{subsec:equilibrium}

The market clears via a standard call auction algorithm designed to maximize total trading volume. Let \(B_{s,r}\) and \(A_{s,r}\) denote the sets of all submitted bids (buy orders) and asks (sell orders), respectively. First, the mechanism identifies all unique price points submitted by the agents to form a discrete set of candidate clearing prices, denoted as \(P_{s,r}\):
\[
    P_{s,r} = \{p \mid (p, q) \in B_{s,r} \cup A_{s,r}\}
\]
In words, this equation simply strips away the quantities and collects every individual price at which any agent has offered to buy or sell during the current round. 

Next, the algorithm calculates the total willingness to buy and sell at every candidate price \(p\). The cumulative demand \(Q^B(p)\) and cumulative supply \(Q^A(p)\) are defined as:
\[
    Q^B(p) = \sum_{(p',q) \in B_{s,r}} q \cdot \mathbbm{1}_{\{p' \geq p\}}
\]
\[
    Q^A(p) = \sum_{(p',q) \in A_{s,r}} q \cdot \mathbbm{1}_{\{p' \leq p\}}
\]
where \(\mathbbm{1}\) is the indicator function. Conceptually, the cumulative demand equation calculates the total number of shares buyers are willing to purchase at price \(p\). It does this by summing the quantities of all buy orders that have a bid price \(p'\) that is greater than or equal to \(p\) (since a buyer willing to pay a higher price is certainly willing to pay a lower price). Conversely, the cumulative supply equation calculates the total number of shares sellers are willing to part with at price \(p\) by summing the quantities of all sell orders with an asking price \(p'\) that is less than or equal to \(p\).

Because a transaction requires both a willing buyer and a willing seller, the actual number of shares that can change hands at any given price \(p\) is constrained by the bottleneck between supply and demand. Therefore, the total executable volume \(V(p)\) is simply the minimum of the cumulative demand and cumulative supply at that price:
\[
    V(p) = \min(Q^B(p), Q^A(p))
\]

Finally, the system must select a single price at which all trades for the round will be executed. The market-clearing price \(p^*_{s,r}\) is determined by evaluating the executable volume \(V(p)\) for every candidate price in our set \(P_{s,r}\), and selecting the price that results in the highest possible number of traded shares:
\[
    p^*_{s,r} = \arg\max_{p \in P_{s,r}} V(p)
\]

To further simplify the state space for the LLM agents, the final clearing price is rounded down to the nearest integer. In the event that there is no intersection between bids and asks (i.e., the highest bid is strictly lower than the lowest ask), no transactions occur. In such cases, the market price for that round is reported as the midpoint between the highest bid and the lowest ask to provide a continuous price signal to the agents.

\subsection{Data Generation and Aggregation}
\label{subsec:data_generation}

Our simulation pipeline systematically constructs a comprehensive panel dataset at the agent-round level. This process yields three distinct layers of empirical observations:

\begin{enumerate}
    \item \textbf{Action and Market Data:} We record all submitted limit orders, executed trades, and end-of-round portfolio values, alongside aggregate market metrics such as the realized clearing price \(P_t\), total trading volume, and bid-ask spreads.
    
    \item \textbf{Expectation Data:} We capture the full term structure of the elicited price forecasts. This enables us to compute the forecast error for agent \(i\) in period \(t\) at horizon \(h\) as:
    \begin{equation}
        E^h_{i,t} = P_{t+h} - f^{(t+h)}_{i,t}
    \end{equation}
    These forecast errors serve as the empirical foundation for our subsequent tests of forecast unbiasedness and zero-autocorrelation.
    
    \item \textbf{Textual Reasoning Data:} We extract the raw text logs from the \texttt{PLANS.txt} and \texttt{INSIGHTS.txt} files, alongside the JSON-formatted \texttt{observations\_and\_thoughts} fields. 
\end{enumerate}

By synchronizing these three data layers, our methodology enables us to observe the macroscopic formation of market prices while simultaneously tracing the micro-level causal chain: from historical market data, to internal textual reasoning, to stated price beliefs, and ultimately to executed trading actions.

\section{Individual Trading Behavior}
\label{sec:behaviors}
A central objective of this study is to dissect the behavioral patterns of AI agents within simulated financial environments. Because Large Language Models (LLMs) are pre-trained on vast corpora of real-world, human-generated data---including historical financial narratives, investor sentiment, and documented market reactions---their outputs theoretically function as sophisticated simulations drawn from a comprehensive database of human behavior. This intrinsic reliance on human data naturally motivates a comparative analysis between AI decision-making and established human cognitive biases. In this section, we propose and formally examine several potential behavioral parallels between AI agents and human investors.

\subsection{Disposition Effect}
One of the most robust findings in behavioral finance is the disposition effect—the well-documented tendency for investors to sell assets that have risen in value (“winners”) too early while holding onto assets that have fallen in value (“losers”) for too long (e.g., \cite{shefrin1985disposition}; \cite{odean1998investors}). For a perfectly rational agent, an asset’s purchase price is a sunk cost and should be irrelevant to current trading decisions, which ought to depend solely on fundamentals and expected returns. To test whether AI agents exhibit this human-like friction, we analyze their trading decisions conditional on the unrealized profitability of their inventory. Table \ref{Table_disposition_effect} reports the results.

The dependent variable is a \emph{Sell Dummy} that equals one if, within a given period, an agent’s total submitted sell orders exceed its total submitted buy orders. The key explanatory variable is a \emph{Gain Dummy} that equals one if the end-of-period $t-1$ asset price exceeds the agent’s weighted-average purchase price (WAPP). We construct WAPP as follows: in round one, for the endowed asset, WAPP is set to its fundamental value of 14. In subsequent rounds, whenever the agent purchases additional shares, WAPP is updated as the quantity-weighted average of the prior WAPP and the new purchase price, with weights equal to the numbers of existing and newly acquired shares. Liquidations do not affect WAPP. As shown in Columns (1) and (2) of Table \ref{Table_disposition_effect}, the coefficient on the Gain Dummy is positive and highly statistically significant. This indicates that when an AI agent holds an unrealized gain, its propensity to sell increases significantly, demonstrating a pronounced disposition effect.

Columns (3) and (4) add the agents’ Average Expectation—defined as the mean of their price forecasts across horizons at the start of period t—as a control. This specification yields two insights. First, the disposition effect is robust: even after controlling for the agents’ forward-looking beliefs, the Gain Dummy remains positive and statistically significant. Thus, the tendency to sell winners is not merely a byproduct of rational forecasting or anticipated mean reversion. Second, beliefs independently shape trading: the coefficient on Average Expectation is strongly negative and highly significant, implying that higher expected future prices rationally reduce the likelihood of selling.

Taken together, these findings reveal a dual-channel mechanism in AI trading behavior. On one hand, agents incorporate forward-looking expectations, reducing sell orders when they anticipate price appreciation. On the other hand, their decisions are simultaneously influenced by past experienced events: the historical entry price exerts an independent and persistent pull on actions, separate from market forecasts. In short, AI agents do not evaluate portfolios as purely frictionless, forward-looking optimizers; instead, their trading behavior remains anchored to historical purchase prices, replicating a canonical human cognitive error.

The persistence of this bias across all LLMs in our sample is particularly striking in light of recent evidence from human markets. \cite{ouyang2025fixed} show that the disposition effect is a fixed individual-level trait that persists across market conditions, suggesting it reflects a stable cognitive feature rather than a context-dependent anomaly. Our findings complement theirs by demonstrating that LLM agents, trained on human-generated text, inherit this stable bias and that it survives even in a controlled experimental setting where the fundamental value is constant and publicly known.

\subsection{Expectation Formation: The Anatomy of AI Return Extrapolation}
\label{sec:extrapolation}
Another prevailing pattern documented in the behavioral finance literature is that human investors are inherently extrapolative; they tend to form beliefs about future market returns based disproportionately on recent past returns (e.g., \cite{greenwood2014}; \cite{barberis2015}; \cite{jin2022asset}). To test whether AI agents exhibit an analogous heuristic, we leverage the incentivized price forecasts elicited from the agents at the conclusion of each trading period. Table~\ref{Table_expectation_formation} presents the regression of the agents' expected future returns---across various forecasting horizons (\(t\), \(t+2\), \(t+5\), and \(t+10\))---on the sequence of realized past returns. 

The results provide compelling evidence of extrapolative belief formation among the AI agents. Across all specifications and forecast horizons, the coefficients on recent realized returns are positive and highly statistically significant, primarily at the 1\% level. This indicates that upon observing a recent price increase, AI agents systematically forecast continued price appreciation, effectively disregarding the mean-reverting pull of the asset's constant fundamental value. 

Furthermore, the data reveal two nuanced dimensions of this extrapolative behavior that closely align with human cognitive limitations, the first being memory decay and recency bias. The coefficients on past returns demonstrate a distinct temporal pattern consistent with the memory literature in behavioral economics and with recent models of memory-based belief formation in financial markets (\cite{charles2024memory}). The agents' expectations are most heavily influenced by the returns of the immediate past (specifically periods \(t-2\) to \(t-1\) and \(t-3\) to \(t-2\)), while the predictive power of older returns (such as \(t-4\) to \(t-3\)) diminishes significantly. For instance, in predicting the immediate next period's return (Column 2), the coefficient drops from \(0.212\) for the one-period lag to \(0.103\) for the three-period lag. This signature pattern suggests that AI agents do not weight historical data equally; rather, their ``working memory'' places a disproportionate emphasis on the most salient, recent market movements.

In addition to recency bias, the results document a pronounced horizon amplification effect. The magnitude of extrapolation increases dramatically as the forecasting horizon extends. When forecasting the return for the current period (\(t\)), the coefficient on the most recent return is \(0.121\) (Column 1). However, when forecasting the return over the subsequent 10 periods (\(t+10\)), the coefficient on the same recent return jumps to \(0.608\) (Column 7). This amplification suggests that AI agents do not merely expect short-term momentum to persist; they aggressively project short-term price trends into long-term structural shifts. This specific cognitive error is frequently cited as a primary driver of endogenous market bubbles in human-dominated markets.

In summary, the empirical evidence presented in Table \ref{Table_expectation_formation} confirms that AI agents do not utilize the provided economic fundamentals to anchor their expectations. Instead, guided by the human-centric data on which they were trained, they rely on a backward-looking, trend-chasing heuristic. This extrapolative nature serves as the foundational cognitive mechanism that fuels the mispricing and speculative bubbles observed throughout our simulated markets.

These findings extend recent evidence that LLMs exhibit extrapolation bias in isolation. \cite{chen2024does} document that ChatGPT produces extrapolative and miscalibrated forecasts when prompted with historical stock returns. Our results demonstrate that this bias is not merely an artifact of single-agent prompting: it persists in an equilibrium market setting where agents receive endogenous price feedback from their own collective behavior, and as we show in Section~\ref{sec:equilibrium}, it has direct consequences for aggregate price dynamics.

\subsection{The Connections between Beliefs and Trading: Translating Expectations into Portfolios}

While the previous subsections establish that AI agents form biased expectations, a critical subsequent question is the extent to which these stated beliefs actually drive their trading behavior. In human markets, the translation of beliefs into portfolio allocations is notoriously imperfect. Notably, \cite{giglio2021} document that while human investors' portfolio allocations are positively correlated with their expected returns, the sensitivity is surprisingly low. Human investors exhibit significant inertia; they rarely trade aggressively on their stated beliefs, possibly due to real-world frictions such as transaction costs, psychological procrastination, lack of confidence, or simple inattention. 

To examine this dynamic within our simulated environment, we regress the AI agents' portfolio allocations (and subsequent changes in positions) on their elicited expected returns. The results of this analysis are reported in Table \ref{Table_expectation_trading}. In stark contrast to the low sensitivity observed in human retail investors by \cite{giglio2021}, the coefficients in Table \ref{Table_expectation_trading} reveal a highly significant, strong positive pass-through from AI beliefs to AI trading decisions. When an AI agent forecasts a positive expected return, it aggressively adjusts its portfolio to overweight the asset, demonstrating a much tighter coupling between stated expectations and executed actions than is typically observed in human data.

This divergence from human behavior is highly plausible when considering the underlying architecture of Large Language Models. Unlike human investors, AI agents do not suffer from physical inertia or the logistical frictions of logging into a brokerage account to execute a trade. More fundamentally, LLMs operate as autoregressive text generators. When an agent is prompted to form a belief (its forecast), that generated text immediately becomes part of its active context window. Because the model's subsequent output (its trading decision) is heavily conditioned on its immediate preceding context, the stated belief acts as a highly salient, frictionless prompt that directly dictates the trading action. Consequently, while AI agents inherit human cognitive biases in forming their beliefs (such as extrapolation and the disposition effect), they execute upon those flawed beliefs with the frictionless efficiency of an algorithmic optimizer.

\section{Equilibrium Price Dynamics}
\label{sec:equilibrium}

While the preceding sections establish that individual AI agents exhibit micro-level cognitive biases---such as the disposition effect and return extrapolation---a critical question remains: do these individual heuristics aggregate into macro-level market dynamics that mirror human trading environments? To address this, we shift our focus from individual portfolio decisions to the equilibrium properties of the simulated market.

\begin{figure}[p]
\centering
\caption{Aggregate Market Price Dynamics Across All LLM Types}\label{Figure_bubble_all_models}
\includegraphics[width=\textwidth]{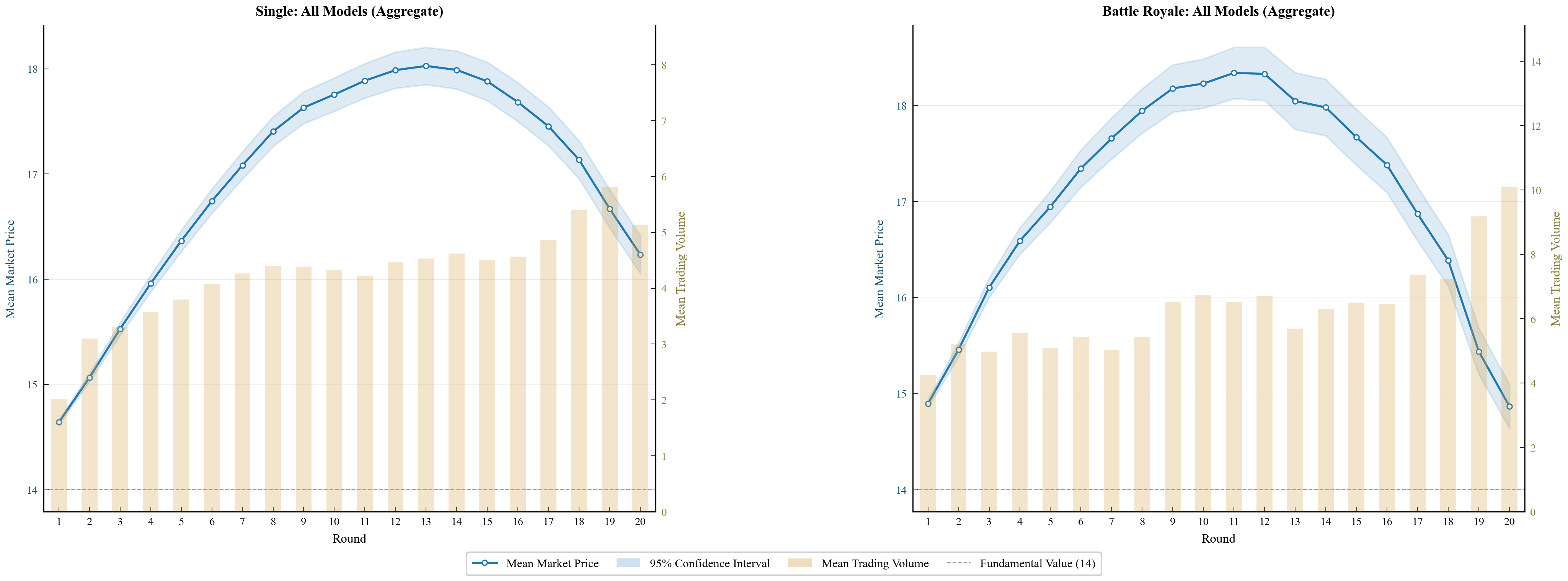}
\vspace{4pt}

{\footnotesize \noindent \textit{Note:} This figure plots the mean market price and trading volume, averaged across all LLM types and simulations, over 20 trading rounds under both single-model and battle-royale market configurations. The solid line traces the cross-simulation mean market price in each round; the shaded band represents the 95\% confidence interval. Bars indicate the mean trading volume per round. The dashed horizontal line marks the fundamental value of 14. In both settings, prices rise above fundamental value in the early rounds and display a hump-shaped trajectory, consistent with bubble-and-crash dynamics commonly documented in experimental asset markets. Trading volume tends to peak around the same rounds as the price apex, suggesting that elevated turnover accompanies mispricing rather than facilitating price correction.}
\end{figure}

Figure~\ref{Figure_bubble_all_models} provides a first answer. Averaging across all LLM types and simulations, mean market prices rise above the fundamental value of 14 in the early trading rounds and trace a hump-shaped trajectory before reverting, a pattern consistent with the bubble-and-crash dynamics documented in human experimental asset markets (\cite{smith1988bubbles}; \cite{palan2013review}). Trading volume peaks near the price apex, suggesting that elevated turnover accompanies mispricing rather than correcting it. Table~\ref{Table_bubble_stat} reveals substantial heterogeneity across model families: some LLMs (e.g., Meta Llama 3.1.70B Instruct, GPT-4o Mini) generate severe mispricing with MSE(FV) exceeding 100, while others (e.g., Deepseek V3, Qwen 2.5) produce near-rational pricing with MSE close to zero. This heterogeneity indicates that the propensity to form bubbles is not a universal artifact of all LLMs but depends on model architecture and training.

The cross-model variation in Table~\ref{Table_bubble_stat} deserves further comment, as it speaks to the robustness and generality of our findings. In Single-Model Markets, MSE(FV) ranges from near zero (Deepseek V3: 0.001; Llama 3.3.70B: 0.05; Qwen 2.5: 0.35) to above 100 (Meta Llama 3.1.70B: 113.8; GPT-4o Mini: 106.0). Even within the same model family the spread can be dramatic: Meta Llama 3.1.70B generates severe bubbles while the newer Llama 3.3.70B prices almost exactly at fundamental value, suggesting that architectural updates and instruction tuning can profoundly alter market-level behavior. A similar within-family divergence appears for Gemini (2.0 Flash: 49.7 vs.\ 2.5 Flash Lite: 0.25). Importantly, the core behavioral regularities we document (extrapolative expectations, the disposition effect, and belief-action coupling) appear across the full spectrum of bubble severity; they are not driven by a single outlier model. The prompt intervention results in Section~\ref{sec:exogenous_manipulation} further confirm generality, as both amplification and suppression effects operate on the pooled sample spanning multiple LLM families.

The Mixed-Model (``Battle Royale'') markets provide additional insight into how interaction between architecturally diverse agents shapes equilibrium outcomes. As Table~\ref{Table_bubble_stat} shows, several mixed pairings produce mispricing levels that lie between those of the constituent single-model markets, consistent with the intuition that pairing a bubble-prone model with a more rational one partially dampens speculative dynamics. For example, the Deepseek V3 $\times$ GPT-4o Mini mixed market generates an MSE(FV) well below that of GPT-4o Mini alone. However, dampening is not universal: some mixed markets exhibit mispricing comparable to the more bubble-prone constituent, suggesting that speculative agents can pull rational counterparts into overvaluation. These patterns indicate that market stability in heterogeneous AI ecosystems will depend not only on the behavioral profiles of individual models but also on the composition and relative weight of different agent types within the trading population.

\subsection{Trajectory of Price Adjustments}
\label{sec:prc_adj}
In classic experimental asset markets, prices rarely jump instantaneously to the fundamental rational expectations equilibrium. Instead, human markets typically exhibit an adaptive, ``Walrasian'' adjustment process. As famously documented by \cite{smith1988bubbles}, the excess of bids over offers (excess demand) in a given trading period serves as a powerful leading indicator for future price movements. In human experiments, a swelling bid-offer gap reliably presages price increases, while a thinning of bids relative to offers signals a loss of behavioral momentum and an impending price drop. 

To test whether our AI-driven markets generate these same endogenous equilibrium dynamics, we analyze the relationship between the bid-offer gap and subsequent price changes:
\begin{equation}
\label{eq:SSW_model}
P_{t} - P_{t-1} = \alpha + \beta (B_{t-1} - O_{t-1}) + \epsilon_t,
\end{equation}
signifies that unrealized trading pressure reflects latent capital gains expectations, which are subsequently realized in market prices. A significant coefficient $\beta > 0$ is therefore diagnostic of a market operating under adaptive, non-equilibrium dynamics rather than instantaneously converging to a rational expectations equilibrium.

Table~\ref{Table_bidoffer_pricechange} presents the results of regressing the one-period-ahead price change on the contemporaneous bid-offer gap, a measure analogous to the excess bids variable. The results are unequivocal: across all specifications, the coefficient on the bid-offer gap is positive and highly statistically significant. This finding {directly replicates the core empirical result of \citet{smith1988bubbles}}. It indicates that in periods where buying interest (bids) outstrips selling interest (offers), prices systematically rise in the subsequent period, and vice versa.

This result carries significant implications for understanding the equilibrium properties of AI markets. It demonstrates that the aggregate outcome of AI agent interaction is not a frictionless, efficient market but one characterized by a {dynamic, expectation-driven feedback loop}. Prices adjust gradually through a process where order flow imbalances---themselves aggregates of individual forecasts---become self-fulfilling prophecies, pushing prices away from or toward fundamental value. This mechanism is precisely what fuels the endogenous boom-and-bust cycles observed in experimental and real-world markets.

Furthermore, this macro-dynamic is logically consistent with and reinforced by the micro-behaviors documented earlier. The {extrapolative expectations} shown in Table~\ref{Table_expectation_formation} provide the cognitive fuel: when agents observe a price increase, they forecast further appreciation, leading them to submit more buy orders, which creates a positive bid-offer gap. The {strong belief-action coupling} evidenced in Table~\ref{Table_expectation_trading} ensures this sentiment is efficiently translated into trading pressure. The predictive power of the bid-offer gap (Table~\ref{Table_bidoffer_pricechange}) then closes the loop, as this pressure materializes into the next period's price increase, thereby validating and reinforcing the initial extrapolative beliefs. This creates a powerful feedback mechanism capable of generating sustained bubbles.

\subsection{Disagreement and Trading Volume: An Equilibrium Link}
\label{subsec:disagreement_volume}

Beyond the dynamic price adjustment process, a second fundamental equilibrium property linking the micro-structure of beliefs to macro-level market outcomes is the relationship between {investor disagreement} and {trading volume}. A core insight from theoretical and empirical finance is that trade requires two parties to hold divergent valuations of an asset. In the absence of liquidity or hedging motives, the primary driver of trading volume in speculative markets is disagreement among investors about the asset's fundamental value \cite{hong2007disagreement}. This theoretical link is a central pillar in understanding the joint dynamics of prices and volume, distinguishing it from models where consensus beliefs prevail.

\cite{hong2007disagreement} emphasize that disagreement---the cross-sectional dispersion in investors' expectations about future cash flows or discount rates---is a necessary condition for trade. In their framework, heterogeneous beliefs generate excess, speculative trading volume. This relationship is not merely a static one; the evolution of disagreement, often fueled by the gradual diffusion of information or differences in opinion formation, can drive both trading activity and price momentum. Our AI-driven markets provide a unique laboratory to test whether this canonical micro-to-macro link emerges naturally from the interaction of AI agents, whose internal beliefs we can directly measure.

We test this hypothesis by examining the predictive power of belief dispersion on aggregate trading volume. Specifically, we construct a measure of {Disagreement in Average Expectation}, defined as the cross-sectional standard deviation of all agents' average price forecasts (across the four horizons) at the beginning of period $t$. We then regress the dollar trading volume in period $t$ on this disagreement measure, controlling for round, simulation, and market type fixed effects.

The results, presented in Table~\ref{Table_belief_disagreement_volume}, confirm a strong, positive relationship. The coefficient on {Disagreement in Average Expectation} is positive and statistically significant across all specifications. For example, in Column (4), which includes the full set of fixed effects, a one-unit increase in the cross-sectional standard deviation of average expectations is associated with an increase in trading volume of approximately 216.8 units ($t = 2.77$). This finding {directly corroborates the central mechanism proposed by \citet{hong2007disagreement}}. It demonstrates that in AI markets, as in human markets, divergent expectations are the fuel for trade. Periods of high disagreement, where agents hold widely varying beliefs about future prices, are systematically associated with elevated trading activity.

This result carries important implications for the equilibrium properties of AI-populated markets. First, it provides a \textit{micro-foundation} for the observed trading volume. The volume is not random noise but is structurally linked to a measurable feature of the agents' internal cognitive states---their dispersed beliefs. Second, it highlights a self-reinforcing dynamic that can contribute to bubble formation and volatility. The extrapolative expectations documented in Section~\ref{sec:extrapolation} can lead to a widening dispersion of forecasts, as agents with different memory weights or initial positions form increasingly divergent views. This rising disagreement, in turn, generates the high trading volume that often characterizes speculative manias, creating a feedback loop between belief divergence, trade, and price discovery.

In summary, the robust link between disagreement and trading volume establishes another profound parallel between AI and human markets at the equilibrium level. It shows that the market microstructure emerging from AI interactions is not only characterized by adaptive price dynamics (Section~\ref{sec:prc_adj}) but also by a volume process that is rationally grounded in the heterogeneity of investor expectations. This finding strengthens the external validity of using AI agents as a model of financial markets, as they replicate a second key equilibrium relationship that is both theoretically predicted and empirically observed in human-dominated markets.

Taken together, the results in this section establish that LLM agents do not merely mimic individual human biases in isolation; they reproduce the aggregate market phenomena that emerge when many biased agents interact within a price-clearing mechanism. The predictive power of excess demand for future prices and the positive disagreement-volume link are equilibrium-level regularities that require the joint operation of extrapolative expectations, belief heterogeneity, and active trading. The fact that these regularities arise endogenously from LLM interactions, without being hard-coded into agent behavior, supports the use of LLM-populated markets as a credible experimental laboratory for studying the formation and collapse of asset price bubbles (we discuss the training data contamination concern in Section \ref{sec:discussion}).

\section{Cognitive Mechanisms}
\label{sec:reasoning}

A persistent challenge in traditional experimental and behavioral finance is the unobservability of the investor's internal cognitive process. While researchers can observe the information set provided to human subjects and their subsequent trading actions, the internal mapping between the two---the actual reasoning process---remains a ``black box.'' Researchers must typically infer these mechanisms from aggregate outcomes or rely on ex-post surveys, which are often noisy. 

In contrast, our LLM-based AI agents provide a unique methodological advantage: their decision-making processes are explicitly generated as text. Through their self-revealed reasoning (captured in their \texttt{PLANS.txt} and \texttt{INSIGHTS.txt} memory files), we can directly observe how agents parse market histories, form expectations, and translate those expectations into concrete trading strategies. We aim to leverage this textual reasoning to unpack the underlying mechanisms of the behavioral findings documented in previous sections. 

Before we can utilize the agents' textual outputs to explain macro-level market anomalies, we must first establish a foundational validation. We must verify that the agents' generated text is not mere ``cheap talk'' or stochastic hallucination, but rather a faithful representation of the cognitive process that governs their quantitative beliefs and actions.

\subsection{Validating the Link: From Reasoning to Beliefs and Trading}

To validate the integrity of the agents' reasoning, we utilize natural language processing to extract the directional intent from the agents' self-revealed text prior to each decision. Specifically, we construct a variable representing the ``Stated Action'' in both their Plans and Insights. This variable takes a value of 1 if the net count of buy-related keywords is positive, -1 if the net count of sell-related keywords is positive, and 0 otherwise. We then test whether these extracted reasoning variables systematically predict the agents' subsequent quantitative forecasts and trading orders.

Table \ref{Table_plan_insights_expectations} investigates the mapping from reasoning to beliefs. Here, we regress the agents' stated expectations of future price movements (their quantitative forecasts for horizons \(t+1\) and \(t+10\)) on the directional intent extracted from their immediately preceding reasoning text. The results confirm a strong, structural link. When an agent's internal monologue articulates a rationale for buying, their subsequent quantitative price forecast is significantly higher. For instance, as shown in Column 1 and Column 3, the coefficients on ``Stated Action in Plans'' and ``Stated Action in Insights'' are positive (\(0.404\) and \(0.855\), respectively) and highly statistically significant. This predictive power holds even for long-term expectations (Columns 5 through 8), confirming that the text accurately captures the formation of the agent's subjective beliefs.

Having established that reasoning drives beliefs, Table \ref{Table_plan_insights_trading} takes the necessary next step: linking reasoning directly to actual trading decisions. A belief is only economically meaningful if it translates into market action. In Table \ref{Table_plan_insights_trading}, we regress the agents' actual order submissions---specifically, a Buy Dummy indicating whether the agent was a net buyer in period \(t+1\)---on their preceding reasoning outputs. 

The empirical estimates in Table \ref{Table_plan_insights_trading} demonstrate a robust translation of thought into action. Agents whose reasoning text reveals a net-buy intent are significantly more likely to submit actual buy orders. The coefficients for both ``Stated Actions in Plan'' and ``Stated Actions in Insights'' are positive (around \(0.205\) and \(0.167\), respectively) and highly significant (t-statistics exceeding 37). This indicates that the text is a primary, causal driver of the agents' interaction with the market clearing mechanism.

Collectively, the results from Tables \ref{Table_plan_insights_expectations} and \ref{Table_plan_insights_trading} provide a crucial methodological validation. They demonstrate that the self-revealed reasoning of AI agents is internally consistent with their stated beliefs and directly translates to their trading behavior. With this validation secured, we can confidently utilize the agents' textual outputs to dissect the precise cognitive heuristics and biases that generate the market dynamics observed in our simulations.

\subsection{Dissecting the Reasoning of AI Agents}

Having established that the self-revealed reasoning of AI agents effectively dictates their quantitative beliefs and trading actions, we now leverage this rich textual data to diagnose the underlying micro-mechanisms driving macro-level market bubbles. The ability to audit the internal reasoning of AI agents provides a unique methodological window into the "black box" of market psychology, allowing us to move beyond correlative evidence to identify the specific heuristics and narratives that aggregate into speculative dynamics.

To systematically analyze the reasoning, we employ an ``LLM-as-a-judge'' methodology. Specifically, we utilize DeepSeek to diagnose the revealed reasoning of the agents. We design a comprehensive prompt that defines twenty distinct behavioral and non-behavioral mechanisms historically linked to bubble formation.\footnote{Appendix section \ref{sec:appendix_audit} provides detailed information about the prompt.}
Table~\ref{Table_mechanism_taxonomy} presents the full taxonomy. The twenty mechanisms are organized into eight categories that span the major theoretical families in the bubble literature: rational bubble theories, extrapolation and expectations, trading biases, confidence and attribution biases, social and herding dynamics, heuristics, risk perception, and narrative and sentiment. Each category is grounded in canonical models of asset pricing and investor behavior, from the rational speculation framework of \cite{blanchard1982bubbles} to the narrative economics of \cite{shiller2017narrative}. This taxonomy provides a systematic and replicable framework for auditing AI reasoning for behavioral content.

For each agent-round observation, the auditing model evaluates the agent's text against these mechanisms and assigns a continuous score ranging from 0 to 1, capturing the relevance and intensity of a focal reasoning feature.

With this dataset, we can directly test which cognitive patterns are most strongly associated with bubble episodes. Our hypothesis is straightforward: if a particular reasoning feature (e.g., momentum-chasing or awareness of a speculative bubble) is a genuine driver of price inflation, then agents' reasoning should exhibit a stronger presence of that feature during bubble periods compared to non-bubble periods. We define a bubble episode as any period where the market price exceeds the fundamental value of 14. We then compare the average numeric score for each of the 20 features between bubble and non-bubble episodes, controlling for investor, round, and simulation fixed effects.

Table \ref{Table_bubble_diff_features} presents our findings. The empirical results reveal a striking anatomy of AI-driven bubbles, characterized primarily by conscious speculation and momentum chasing. The largest positive differences are observed in the \textit{Rational Speculative Bubble} and \textit{Momentum vs. Newswatcher} features. For instance, the propensity to engage in momentum trading over fundamental news-watching increases significantly during bubbles (Difference = \(0.181\) in Insights, \(t = 32.135\)). Similarly, agents explicitly exhibit reasoning consistent with a \textit{Rational Speculative Bubble}, acknowledging the overvaluation but participating with the intent to sell to future buyers (Difference = \(0.213\) in Insights, \(t = 27.679\)). Concurrently, we observe a significant abandonment of fundamental anchors. The coefficient for \textit{Extrapolation vs. Anchor} is significantly negative (Difference = \(-0.073\) in Plans, \(t = -10.695\)); because a lower score on this metric corresponds to pure extrapolation rather than fundamental anchoring, this negative gap confirms that agents actively untether their expectations from the asset's intrinsic value during explosive price rallies.

Furthermore, the analysis highlights the exacerbating role of coordination frictions and cognitive biases. During bubble episodes, agents demonstrate a heightened awareness of \textit{Synchronization Risk} (Difference = \(0.047\), \(t = 12.594\)), explicitly recognizing the coordination problem inherent in timing their market exit. Yet, their inherent \textit{Disposition Effect} intensifies (Difference = \(0.060\), \(t = 14.483\)), further distorting the supply-demand equilibrium as they navigate their unrealized gains and losses. Ultimately, this textual dissection confirms that AI-driven bubbles do not emerge from random noise or mere algorithmic error. Instead, they are fueled by agents actively articulating and executing classic behavioral and speculative mechanisms, which collectively aggregate into severe market mispricing.

In summary, the self-revealed reasoning shows that AI agents, much like human investors during historical bubbles, are often not ignorant of the overvaluation; rather, they knowingly participate in the bubble, betting on their ability to time the market. This finding provides novel, direct evidence of the internal narratives that underlie the equilibrium dynamics documented earlier, closing the loop from micro-level cognition to macro-level market phenomena.

\section{Policy Intervention via Prompt Design}
\label{sec:exogenous_manipulation}

Having identified the significant distinction in heuristics during bubble and non-bubble episodes, we now propose how strategically manipulating the tendency of AI agents' heuristics can help affect the market equilibrium. To achieve this, we introduce exogenous shocks via prompts to the agents' cognitive processes. Concretely, we design exogenous “instructional shocks” embedded in AI agent's context that either amplify or suppress targeted heuristics for every agent in the market and study how these manipulations propagate from micro-level cognition to macro-level prices and bubble magnitudes. Specifically, we consider two types of shocks:
\begin{itemize}
    \item \textbf{Amplification shocks:} Prompts explicitly encourage the targeted heuristic (e.g., momentum-chasing over fundamentals, ``ride the bubble'' resale logic, new-era narratives). The aim is to strengthen tendencies that Table \ref{Table_bubble_diff_features} showed to be elevated during bubbles.
    \item \textbf{Suppression shocks:} Prompts explicitly discourage the targeted heuristic and redirect attention toward fundamentals, statistical discipline, and risk control. For example, the ``Momentum vs.\ Newswatcher'' suppression shock instructs agents to downweight recent price trends and prioritize the constant fundamental value, documented cash flows, and formal inference.
\end{itemize}
For a given type of shock, we implement it market-wide to every AI agent. 

The empirical results in Table \ref{Table_amp_sup_stat} demonstrate that AI agents are highly responsive to exogenous cognitive shocks, allowing us to experimentally manipulate market dynamics. Under the Amplification treatment, we observe massive, statistically significant surges in bubble-driving heuristics. Most notably, the \textit{Rational Speculative Bubble} feature increases by 5.07 (\(t = 5.43\)) relative to the benchmark, while \textit{Momentum vs.\ Newswatcher} and \textit{New Era Thinking} spike by 5.32 (\(t = 5.97\)) and 5.57 (\(t = 4.81\)), respectively. By exogenously feeding the agents' latent biases, we successfully engineer an environment where agents aggressively abandon fundamental anchors in favor of pure trend-following, thereby exacerbating the magnitude of market bubbles. 

Conversely---and more importantly for market design---the Suppression treatment proves highly effective at ``debiasing'' the agents. When provided with rational guardrails, agents significantly reduce their reliance on flawed heuristics. For instance, the \textit{Extrapolation vs.\ Anchor} score drops by 2.23 (\(t = -6.24\)) compared to the benchmark, indicating that agents successfully shift their focus away from past price trajectories and back toward the asset's intrinsic fundamental value. We also observe significant reductions in \textit{Diagnostic Expectations} (Difference = -2.00, \(t = -3.32\)) and \textit{Narrative Tone} (Difference = -1.97, \(t = -3.12\)). By dampening these specific cognitive errors, the suppression prompts effectively cool down the speculative fervor, resulting in market prices that are much more tightly bound to the fundamental value (as reflected in the Sup.\ vs.\ FV column).

These findings carry profound fundamental implications for regulatory policy in an era increasingly dominated by algorithmic and AI-driven trading. In traditional human-dominated markets, regulators struggle to eliminate endogenous bubbles because human psychology and cognitive biases are notoriously difficult to ``patch'' or rewire. However, our results reveal that AI agents are fundamentally different: their cognitive frameworks are entirely programmable.

Our debiasing result also speaks to the broader literature on LLM rationality. Recent work has explored two potential remedies for AI behavioral biases: scaling model size (\cite{bini2025behavioral}) and increasing reasoning effort (\cite{cartea2026aibubbles}). Bini et al. find that larger models exhibit fewer biases, while Cartea et al. show that higher reasoning effort reduces but does not eliminate speculative participation. Our prompt intervention approach offers a complementary and more targeted lever: rather than relying on ``think harder'' (more compute) or ``think longer'' (more reasoning steps), behavioral-finance-informed prompts direct agents to ``think differently'' by suppressing specific cognitive errors. The effectiveness of this approach suggests that the path to stable AI-driven markets lies not only in model architecture but also in the design of cognitive guardrails grounded in established behavioral theory. 

By ``educating'' these AI agents through carefully engineered system prompts, we can systematically reduce market inefficiency. This suggests a novel, highly actionable paradigm for financial regulation. Rather than relying solely on post-trade circuit breakers or capital requirements, regulators could mandate specific ``cognitive guardrails'' or debiasing instructions within the system prompts of commercially deployed trading AIs. By hardcoding fundamental anchoring and suppressing extrapolative heuristics at the prompt level, policymakers possess a direct, ex-ante mechanism to foster systemic market stability and prevent the formation of AI-driven speculative crashes.

\section{Discussion}
\label{sec:discussion}

Before concluding, we address several considerations that constrain the interpretation of our findings.

\subsection{Training Data and Emergent Behavior}
\label{subsec:contamination}

A central interpretive question is whether the behavioral biases we document reflect emergent properties of multi-agent interaction or merely the reproduction of patterns from the models' training corpora. LLMs are trained on text that includes finance textbooks, descriptions of behavioral biases, and even accounts of the SSW experimental paradigm. We do not claim to definitively resolve whether these biases reflect emergent cognition or sophisticated pattern reproduction, a question that may ultimately require advances in mechanistic interpretability research. However, three features of our results constrain the set of viable explanations and suggest that the behavioral patterns are not merely superficial mimicry.

First, the equilibrium-level regularities we document---the predictive power of excess demand for future prices, the positive disagreement-volume relationship---emerge from the \textit{interaction} of multiple agents within a price-clearing mechanism, not from any single agent's text generation. These aggregate patterns cannot be hard-coded into individual agent outputs; they require the joint operation of heterogeneous beliefs, active trading, and endogenous price feedback. Second, the prompt intervention results in Section~\ref{sec:exogenous_manipulation} demonstrate that agent behavior responds causally and directionally to exogenous cognitive shocks. When we amplify momentum-chasing heuristics, bubble magnitudes increase; when we suppress them, bubbles shrink. This systematic responsiveness to targeted interventions is difficult to reconcile with rote pattern matching. Third, the marked heterogeneity in bubble magnitude across model families---from near-zero mispricing (Deepseek V3) to severe bubbles (Meta Llama 3.1.70B)---suggests that behavioral outcomes depend on model architecture and training composition rather than on a universal template that all LLMs reproduce identically.

Regardless of the ultimate mechanistic explanation, the market consequences are the same: whether an LLM ``truly'' has a disposition effect or merely reproduces one with high fidelity, the resulting mispricing, volume dynamics, and bubble trajectories are real features of the market equilibrium. From a financial stability perspective, the distinction between emergent and reproduced biases is less important than the fact that these biases are systematic, predictable, and amenable to intervention.

\subsection{Scope and Limitations}
\label{subsec:limitations}

Several limitations of our design deserve mention. First, our markets are populated exclusively by AI agents; real-world markets feature both human and AI participants whose interactions may produce qualitatively different dynamics. Second, our agents operate within a single experimental paradigm (the SSW open-call auction with constant fundamental value), and the degree to which our findings generalize to richer institutional settings---including continuous double auctions, derivative markets, or environments with time-varying fundamentals---remains an open question. Third, commercially deployed trading systems often rely on reinforcement learning or fine-tuned models rather than raw prompt-driven LLMs, so our policy implications regarding prompt-level cognitive guardrails are most directly applicable to the growing class of LLM-based trading tools and advisory systems where prompt-level interventions are feasible. Finally, the rapid pace of model development means that behavioral profiles may shift as new architectures and training procedures emerge.

\section{Conclusion}
\label{sec:conclusion}
This paper provides comprehensive evidence that autonomous LLM agents, when deployed in an equilibrium asset market, systematically reproduce the behavioral patterns documented in decades of human investor research, and that these tendencies have measurable consequences for market stability. By deploying fourteen frontier models within a standard Walrasian open-call auction, we trace the full causal chain from individual cognitive patterns through trading behavior to aggregate price dynamics.

Our findings reveal that AI traders, trained on vast corpora of human-generated data, inherently adopt the behavioral heuristics and cognitive patterns of human investors. At the micro-level, we document that AI agents systematically exhibit the disposition effect and form heavily extrapolative beliefs. However, unlike human retail investors—whose stated beliefs often show a weak correlation with actual portfolio allocations due to physical and psychological frictions (\cite{giglio2021})—AI agents exhibit a remarkably strong, frictionless coupling between their expectations and trading actions. This unique characteristic allows us to observe the unattenuated impact of these behavioral tendencies on market mechanics.

At the equilibrium level, we show that these individual patterns organically generate complex, well-documented market phenomena. Cross-sectional disagreement among AI agents strongly correlates with trading volume, and the bid-offer gap (excess demand) reliably predicts future price trajectories, successfully replicating the classic experimental findings of \cite{smith1988bubbles}. Furthermore, by leveraging the unique ability to extract the AI agents' "inner monologues," we provide direct textual evidence that these market dynamics are driven by explicit behavioral reasoning. During market bubbles, agents consciously pivot toward speculative, momentum-chasing strategies, whereas non-bubble periods are characterized by fundamental valuation reasoning.

Crucially, the very features that make AI agents susceptible to these behavioral patterns—their reliance on prompt-driven reasoning and their strong belief-action coupling—also make those tendencies amenable to intervention. We demonstrate that the cognitive patterns of AI agents are highly programmable. By introducing behavioral-finance-informed "recalibrating" prompts, we successfully mitigate extrapolative behaviors and significantly reduce the magnitude of market bubbles. Conversely, prompts designed to amplify these tendencies predictably exacerbate bubble formation, confirming the causal link between agent heuristics and market instability.

Our results complement concurrent findings that LLMs trade more rationally than humans in controlled settings (\cite{henning2025llm}) and that increased reasoning effort alone does not eliminate speculative bubbles (\cite{cartea2026aibubbles}). Where those studies identify the problem, we offer a solution: behavioral-finance-informed prompt interventions provide an actionable mechanism for stabilizing AI-driven markets. As financial institutions continue to delegate complex pricing and allocation tasks to AI, systemic stability will depend not only on traditional market regulations but also on the implementation of cognitive guardrails at the prompt level. Important open questions remain: whether these patterns and interventions generalize to heterogeneous markets with both AI and human traders, whether inter-agent learning can erode or entrench behavioral tendencies over time, and how rapidly evolving model architectures will reshape the landscape of AI behavioral finance.







\clearpage
\bibliography{References}

\clearpage

\clearpage

\clearpage
\begin{table}[p]
\caption{Market Performance Metrics Across LLM Types}\label{Table_bubble_stat}
{This table summarises two key metrics of market quality for each LLM type under both single-model and battle-royale market configurations. \textbf{MSE (FV)} is the mean squared error of observed market prices relative to the fundamental value of 14, measuring the magnitude of mispricing across rounds and simulations. \textbf{PV Var.}\ is the average within-period cross-agent variance in portfolio value, capturing the degree of strategy heterogeneity among traders. Higher MSE values indicate larger sustained deviations from rational pricing, while higher PV Variance reflects greater dispersion in individual trading outcomes.}\\
\begin{center}
\begin{tabular}{l cc cc}
\toprule
 & \multicolumn{2}{c}{Single Mode} & \multicolumn{2}{c}{Battle Royale Mode} \\
\cline{2-3}\cline{4-5}
 & MSE (FV) & PV Var. & MSE (FV) & PV Var. \\
\midrule
Claude 3.5 Sonnet & 18.4625 & 68.3700 & 2.4125 & 8.4100 \\
Claude 3.7 Sonnet & 27.3275 & 372.18 & 29.9181 & 253.86 \\
Cogito V2 Preview Deepseek 671b & 0.2500 & 6.9300 & 0.3203 & 2.4300 \\
Deepseek V3.0324 & 0.0012 & 0.0000 & 2.0440 & 8.3300 \\
Gemini 2.0 Flash 001 & 49.6888 & 119.66 & 33.0219 & 223.55 \\
Gemini 2.5 Flash Lite & 0.2512 & 3.7900 & 3.7017 & 435.73 \\
Gpt 4o & 27.6005 & 64.1700 & 21.8449 & 105.19 \\
Gpt 4o Mini & 105.99 & 278.11 & 53.2375 & 109.28 \\
Grok 3 & 3.0888 & 73.0700 & 5.9188 & 51.9600 \\
Llama 3.3.70B Instruct & 0.0500 & 0.0000 & 8.2875 & 0.0200 \\
Llama 4 Maverick 17B 128E Instruct Fp8 & 8.6227 & 50.2600 & 15.4266 & 7.5200 \\
Meta Llama 3.1.70B Instruct & 113.83 & 31.2000 & 99.8125 & 32.7500 \\
Qwen2.5.72b Instruct & 0.3534 & 5.0600 & 0.0000 & 0.0000 \\
Qwen3.235b A22b Instruct 2507 & 0.2625 & 20.6300 & 1.5062 & 4.7100 \\
\bottomrule
\end{tabular}
\end{center}
\end{table}
\clearpage

\begin{landscape}
\begin{table}[p]
\caption{The Disposition Effect}\label{Table_disposition_effect}
{This table first replicates the disposition effect observed in human investors and then compares it to the influence of expectations on the selling decisions of AI agents. The dependent variable is a \textit{Sell Dummy}, which equals one if the total number of submitted selling orders exceeds the total number of submitted buying orders. In columns (1) and (2), the key explanatory variable, \textit{Gain Dummy}, equals one if the asset price at the end of period $t-1$ is higher than the weighted average purchase price of the focal asset, and zero otherwise. In columns (3) and (4), we control for investors' average expectations, measured as the mean of their forecasts across different horizons at the beginning of period $t$. The analysis includes investor fixed effects, round fixed effects, and simulation fixed effects. Standard errors are clustered at the AI agent level.}\\
\begin{center}
{
\def\sym#1{\ifmmode^{#1}\else\(^{#1}\)\fi}
\begin{tabular*}{1.0\hsize}{@{\hskip\tabcolsep\extracolsep\fill}l*{4}{cc}}
\toprule
                    &\multicolumn{4}{c}{Sell Dummy}\\
                    \cline{2-5}
                    &\multicolumn{1}{c}{(1)}&\multicolumn{1}{c}{(2)}&\multicolumn{1}{c}{(3)}&\multicolumn{1}{c}{(4)}\\
\midrule
Gain Dummy          &       0.050\sym{***}&       0.039\sym{***}&       0.016\sym{*}  &       0.033\sym{***}\\
                    &      (6.68)         &      (4.36)         &      (2.07)         &      (3.70)         \\
Average Expectation &                     &                     &      -1.122\sym{***}&      -0.606\sym{***}\\
                    &                     &                     &    (-26.71)         &    (-14.06)         \\
\midrule
Investor FE         &         YES         &         YES         &         YES         &         YES         \\
Round FE            &          NO         &         YES         &          NO         &         YES         \\
Simulation FE       &          NO         &         YES         &          NO         &         YES         \\
Adjusted R-Squared  &       0.238         &       0.331         &       0.307         &       0.354         \\
N                   &      65,303         &      65,303         &      60,850         &      60,850         \\

\bottomrule
\end{tabular*}
}

\end{center}
\end{table}
\end{landscape}
\clearpage

\begin{landscape}
\begin{table}[!htb]
\caption{Expectation Formation: Extrapolating Past Returns}\label{Table_expectation_formation}
{This table examines the predictive power of past price changes on LLM-agent expectations. The dependent variable is the agent's forecast made at the end of period t. The column headers indicate the forecast horizon, where $Expectation[R_{t+N}]$ represents expected returns from the beginning of period t+1 through the end of period t+N. The primary explanatory variables are realized past returns. All specifications include investor, round, and simulation fixed effects, with standard errors clustered at the AI agent level. }\\
\begin{center}
{
\def\sym#1{\ifmmode^{#1}\else\(^{#1}\)\fi}
\begin{tabular*}{1.0\hsize}{@{\hskip\tabcolsep\extracolsep\fill}l*{8}{cc}}
\toprule
                    &\multicolumn{2}{c}{Expectation[$R_{t}$]} &\multicolumn{2}{c}{Expectation[$R_{t+2}$]}  &\multicolumn{2}{c}{Expectation[$R_{t+5}$]} &\multicolumn{2}{c}{Expectation[$R_{t+10}$]} \\ 
                    \cline{2-3}\cline{4-5}\cline{6-7}\cline{8-9}
                    &\multicolumn{1}{c}{(1)}&\multicolumn{1}{c}{(2)}&\multicolumn{1}{c}{(3)}&\multicolumn{1}{c}{(4)}&\multicolumn{1}{c}{(5)}&\multicolumn{1}{c}{(6)}&\multicolumn{1}{c}{(7)}&\multicolumn{1}{c}{(8)}\\
\midrule
Recent Return (t-1 to t)&       0.121\sym{***}&       0.121\sym{***}&       0.392\sym{***}&       0.347\sym{***}&       0.552\sym{***}&       0.468\sym{***}&       0.608\sym{***}&       0.438\sym{***}\\
                    &     (12.29)         &     (12.57)         &     (21.43)         &     (19.82)         &     (21.99)         &     (19.28)         &     (18.78)         &     (13.74)         \\
Recent Return (t-2 to t-1)&       0.219\sym{***}&       0.212\sym{***}&       0.471\sym{***}&       0.413\sym{***}&       0.601\sym{***}&       0.488\sym{***}&       0.565\sym{***}&       0.380\sym{***}\\
                    &     (29.21)         &     (27.30)         &     (34.73)         &     (30.24)         &     (32.10)         &     (25.72)         &     (20.88)         &     (13.69)         \\
Recent Return (t-3 to t-2)&       0.189\sym{***}&       0.182\sym{***}&       0.356\sym{***}&       0.297\sym{***}&       0.434\sym{***}&       0.313\sym{***}&       0.316\sym{***}&       0.120\sym{***}\\
                    &     (29.54)         &     (26.87)         &     (31.65)         &     (25.09)         &     (26.28)         &     (18.08)         &     (12.68)         &      (4.60)         \\
Recent Return (t-4 to t-3)&       0.109\sym{***}&       0.103\sym{***}&       0.212\sym{***}&       0.159\sym{***}&       0.235\sym{***}&       0.128\sym{***}&      0.0841\sym{**} &     -0.0808\sym{*}  \\
                    &     (15.34)         &     (14.11)         &     (15.12)         &     (11.05)         &     (10.79)         &      (5.89)         &      (2.61)         &     (-2.55)         \\
\midrule
Investor FE         &         YES         &         YES         &         YES         &         YES         &         YES         &         YES         &         YES         &         YES         \\
Round FE            &          NO         &         YES         &          NO         &         YES         &          NO         &         YES         &          NO         &         YES         \\
Simulation FE       &          NO         &         YES         &          NO         &         YES         &          NO         &         YES         &          NO         &         YES         \\
Adjusted R-Squared  &       0.509         &       0.510         &       0.598         &       0.610         &       0.645         &       0.665         &       0.666         &       0.686         \\
N                   &      47,733         &      47,733         &      47,725         &      47,725         &      47,719         &      47,719         &      47,719         &      47,719         \\
\bottomrule
\end{tabular*}
}
\end{center}
\end{table}
\end{landscape}
\clearpage

\begin{landscape}
\begin{table}[!htb]
\caption{Expectation and Trading Decisions}\label{Table_expectation_trading}
{This table presents the predictive power of LLM-agent expectations on their trading decisions. The dependent variable is an indicator for buying decisions, which equals one if the LLM-agent net buys during the period $t + 1$, and zero otherwise. The explanatory variables are their forecasts made at the end of period $t$. All specifications include investor, round, and simulation fixed effects, with standard errors clustered at the  AI agent level.}\\
\begin{center}
{
\def\sym#1{\ifmmode^{#1}\else\(^{#1}\)\fi}
\begin{tabular*}{1.0\hsize}{@{\hskip\tabcolsep\extracolsep\fill}l*{8}{cc}}
\toprule
                    &\multicolumn{8}{c}{Buy Dummy}\\
                    \cline{2-9}
                    &\multicolumn{1}{c}{(1)}&\multicolumn{1}{c}{(2)}&\multicolumn{1}{c}{(3)}&\multicolumn{1}{c}{(4)}&\multicolumn{1}{c}{(5)}&\multicolumn{1}{c}{(6)}&\multicolumn{1}{c}{(7)}&\multicolumn{1}{c}{(8)}\\
\midrule
Expectation (t)     &       0.660\sym{***}&       0.504\sym{***}&                     &                     &                     &                     &                     &                     \\ 
                    &      (7.42)         &      (5.87)         &                     &                     &                     &                     &                     &                     \\
Expectation (t+2)   &                     &                     &       0.400\sym{***}&       0.340\sym{***}&                     &                     &                     &                     \\ 
                    &                     &                     &      (8.61)         &      (7.04)         &                     &                     &                     &                     \\
Expectation (t+5)   &                     &                     &                     &                     &       0.294\sym{***}&       0.286\sym{***}&                     &                     \\ 
                    &                     &                     &                     &                     &      (9.33)         &      (8.58)         &                     &                     \\
Expectation (t+10)  &                     &                     &                     &                     &                     &                     &       0.221\sym{***}&       0.220\sym{***}\\ 
                    &                     &                     &                     &                     &                     &                     &      (9.65)         &      (9.28)         \\
\midrule
Investor FE         &         YES         &         YES         &         YES         &         YES         &         YES         &         YES         &         YES         &         YES         \\
Round FE            &          NO         &         YES         &          NO         &         YES         &          NO         &         YES         &          NO         &         YES         \\
Simulation FE       &          NO         &         YES         &          NO         &         YES         &          NO         &         YES         &          NO         &         YES         \\
Adjusted R-Squared  &       0.515         &       0.519         &       0.517         &       0.520         &       0.518         &       0.521         &       0.518         &       0.521         \\
N                   &      63,888         &      63,888         &      63,880         &      63,880         &      63,874         &      63,874         &      63,874         &      63,874         \\
\bottomrule
\end{tabular*}
}

\end{center}
\end{table}
\end{landscape}
\clearpage

\begin{table}[!htb]
\caption{Bid-Offer Gaps and Future Price Changes}\label{Table_bidoffer_pricechange}
{This table presents the predictive power of bid-offer gaps on future price changes. The dependent variable is the future price change, and the explanatory variable is the bid-offer gap. All specifications include round, simulation, and market type fixed effects. Standard errors are clustered at the simulation level.}\\
\begin{center}
{
\def\sym#1{\ifmmode^{#1}\else\(^{#1}\)\fi}
\begin{tabular*}{1.0\hsize}{@{\hskip\tabcolsep\extracolsep\fill}l*{6}{cc}}
\toprule
                    &\multicolumn{6}{c}{Price Changes (t to t+1)} \\
                    \cline{2-7}
                    &\multicolumn{1}{c}{(1)}&\multicolumn{1}{c}{(2)}&\multicolumn{1}{c}{(3)}&\multicolumn{1}{c}{(4)}&\multicolumn{1}{c}{(5)}&\multicolumn{1}{c}{(6)}\\
\midrule
Bid-Offer           &      0.0149\sym{***}&     0.00966\sym{***}&      0.0179\sym{***}&     0.00979\sym{***}&      0.0179\sym{***}&     0.00847\sym{***}\\
                    &      (8.30)         &      (5.26)         &      (7.76)         &      (5.53)         &      (7.75)         &      (3.48)         \\
Intercept            &      0.0327\sym{*}  &      0.0456\sym{**} &      0.0251\sym{***}&      0.0453\sym{**} &      0.0251\sym{***}&      0.0486\sym{***}\\
                    &      (2.22)         &      (3.16)         &      (4.38)         &      (3.15)         &      (4.38)         &      (8.04)         \\
\midrule
Round FE            &          NO         &         YES         &          NO         &         YES         &          NO         &         YES         \\
Simulation FE       &          NO         &          NO         &         YES         &         YES         &         YES         &         YES         \\
Market Type FE      &          NO         &          NO         &          NO         &          NO         &         YES         &         YES         \\
Adjusted R-Squared  &       0.123         &       0.266         &       0.189         &       0.269         &       0.189         &       0.333         \\
N                   &       3,211         &       3,211         &       3,211         &       3,211         &       3,211         &       3,211         \\

\bottomrule
\end{tabular*}
}

\end{center}
\end{table}
\clearpage

\begin{table}[!htb]
\caption{Disagreement in Expectations and Dollar Trading Volume}\label{Table_belief_disagreement_volume}
{This table examines the predictive power of disagreement in expectations for future dollar trading volume. The dependent variable is the dollar trading volume in period t. The key explanatory variable is the cross-sectional standard deviation of AI agents' average expectations (across the four forecasting horizons), measured at the beginning of period t. All regressions include round, simulation, and market type fixed effects. Standard errors are clustered at the simulation level.}\\
\begin{center}
{
\def\sym#1{\ifmmode^{#1}\else\(^{#1}\)\fi}
\begin{tabular*}{1.0\hsize}{@{\hskip\tabcolsep\extracolsep\fill}l*{5}{cc}}
\toprule
                    &\multicolumn{4}{c}{Trading Dollar Volume} \\
                    \cline{2-5}
                    &\multicolumn{1}{c}{(1)}&\multicolumn{1}{c}{(2)}&\multicolumn{1}{c}{(3)}&\multicolumn{1}{c}{(4)}\\
\midrule
Disagreement in Average Expectation&       322.9\sym{**} &       334.3\sym{**} &       177.6\sym{*}  &       216.8\sym{**} \\
                    &      (3.03)         &      (3.15)         &      (2.12)         &      (2.77)         \\
constant            &       60.77\sym{***}&       60.27\sym{***}&       67.11\sym{***}&       65.40\sym{***}\\
                    &      (9.63)         &      (9.57)         &     (18.37)         &     (19.12)         \\
\midrule
Round FE            &          NO         &         YES         &          NO         &         YES         \\
Simulation FE       &          NO         &          NO         &         YES         &         YES         \\
Market Type FE      &          NO         &          NO         &          NO         &         YES         \\
Adjusted R-Squared  &       0.029         &       0.046         &       0.569         &       0.590         \\
N                   &       3,032         &       3,032         &       3,032         &       3,032         \\
\bottomrule
\end{tabular*}
}

\end{center}
\end{table}
\clearpage

\begin{landscape}
\begin{table}[!htb]
\caption{Plan\& Insights and Investor Expectations}\label{Table_plan_insights_expectations}
{This table examines the predictive power of investors' stated actions—derived from their plans and insights—on their return expectations. The dependent variable is the agent's forecast formed at the end of period t. The column headers indicate the forecast horizon, where $Expectation[R_{t+N}]$ represents expected returns from the beginning of period t+1 through the end of period t+N. The primary explanatory variable represents the actions stated in investors' plans and insights as of the end of period t. This variable takes a value of 1 if the net count of buy-related keywords is positive, -1 if the net count of sell-related keywords is positive, and 0 otherwise. All specifications include investor, round, and simulation fixed effects, with standard errors clustered at the simulation level. }\\
\begin{center}
{
\def\sym#1{\ifmmode^{#1}\else\(^{#1}\)\fi}
\begin{tabular*}{1.0\hsize}{@{\hskip\tabcolsep\extracolsep\fill}l*{8}{cc}}
\toprule
                    &\multicolumn{4}{c}{Expectation[$R_{t+1}$]} &\multicolumn{4}{c}{Expectation[$R_{t+10}$]}\\
                    \cline{2-5} \cline{6-9}
                    &\multicolumn{1}{c}{(1)}&\multicolumn{1}{c}{(2)}&\multicolumn{1}{c}{(3)}&\multicolumn{1}{c}{(4)}&\multicolumn{1}{c}{(5)}&\multicolumn{1}{c}{(6)}&\multicolumn{1}{c}{(7)}&\multicolumn{1}{c}{(8)}\\
\midrule
Stated Action in Plans&       0.404\sym{***}&       0.103         &                     &                     &       2.914\sym{***}&       1.672\sym{***}&                     &                     \\
                    &      (5.95)         &      (1.85)         &                     &                     &      (7.80)         &      (5.57)         &                     &                     \\
Stated Action in Insights&                     &                     &       0.855\sym{***}&       0.449\sym{***}&                     &                     &       4.565\sym{***}&       2.805\sym{***}\\
                    &                     &                     &     (13.38)         &      (9.69)         &                     &                     &     (14.03)         &     (11.31)         \\
\midrule
Investor FE         &         YES         &         YES         &         YES         &         YES         &         YES         &         YES         &         YES         &         YES         \\
Round FE            &          NO         &         YES         &          NO         &         YES         &          NO         &         YES         &          NO         &         YES         \\
Simulation FE       &          NO         &         YES         &          NO         &         YES         &          NO         &         YES         &          NO         &         YES         \\
Adjusted R-Squared  &       0.270         &       0.381         &       0.281         &       0.385         &       0.486         &       0.595         &       0.496         &       0.599         \\
N                   &      63,888         &      63,888         &      63,888         &      63,888         &      63,874         &      63,874         &      63,874         &      63,874         \\
\bottomrule
\end{tabular*}
}

\end{center}
\end{table}
\end{landscape}
\clearpage

\begin{landscape}
\begin{table}[!htb]
\caption{Plan\& Insights and Investor Trading Behaviors}\label{Table_plan_insights_trading}
{This table presents the predictive power of investors' stated actions in their plans and insights on their actual trading decisions. The dependent variable is an indicator for buying decisions, which equals one if the LLM-agent net buys during the period $t + 1$, and zero otherwise.  The primary explanatory variable represents the actions stated in investors' plans and insights as of the end of period t. This variable takes a value of 1 if the net count of buy-related keywords is positive, -1 if the net count of sell-related keywords is positive, and 0 otherwise. All specifications include investor, round, and simulation fixed effects, with standard errors clustered at the simulation level. }\\
\begin{center}
{
\def\sym#1{\ifmmode^{#1}\else\(^{#1}\)\fi}
\begin{tabular*}{1.0\hsize}{@{\hskip\tabcolsep\extracolsep\fill}l*{4}{cc}}
\toprule
                    &\multicolumn{4}{c}{Buy Dummy} \\
                    \cline{2-5}
                    &\multicolumn{1}{c}{(1)}&\multicolumn{1}{c}{(2)}&\multicolumn{1}{c}{(3)}&\multicolumn{1}{c}{(4)}\\
\midrule
Stated Actions in Plan&       0.205\sym{***}&       0.202\sym{***}&                     &                     \\
                    &     (71.63)         &     (74.07)         &                     &                     \\
Stated Actions in Insights&                     &                     &       0.167\sym{***}&       0.165\sym{***}\\
                    &                     &                     &     (37.49)         &     (39.42)         \\

\midrule
Investor FE         &         YES         &         YES         &         YES         &         YES         \\
Round FE            &          NO         &         YES         &          NO         &         YES         \\
Simulation FE       &          NO         &         YES         &          NO         &         YES         \\
Adjusted R-Squared  &       0.549         &       0.550         &       0.528         &       0.529         \\
N                   &      68,739         &      68,739         &      68,739         &      68,739         \\
\bottomrule
\end{tabular*}
}

\end{center}
\end{table}
\end{landscape}
\clearpage

\begin{landscape}
\begin{table}[!htb]
\caption{Comparison of AI Reasoning Features During Bubble vs. Non-Bubble Episodes}\label{Table_bubble_diff_features}
{This table compares the strength of various reasoning features during bubble and non-bubble episodes. Specifically, we examine twenty distinct features derived from the reasoning of an AI agent, as reflected in its insights and plans, using Deepseek. The observed differences represent the gap in a focal feature between bubble and non-bubble episodes, where bubble episodes are defined as those with prices exceeding their fundamental value (14). In estimating these differences, we control for investor, round, and simulation fixed effects.}\\
\begin{center}
\begin{tabular}{l*{5}{c}}
\hline
&\multicolumn{2}{c}{Insights} &\multicolumn{2}{c}{Plans} \\
\cline{2-3}\cline{4-5}
Features&\multicolumn{1}{c}{Difference}&\multicolumn{1}{c}{t-value}&\multicolumn{1}{c}{Difference}&\multicolumn{1}{c}{t-value} \\
\toprule
Rational Speculative Bubble&0.213&27.679&0.204&29.190\\
Synchronization Risk&0.047&12.594&0.040&8.699\\
Asymmetric Information&0.004&3.330&0.007&2.766\\
Extrapolation vs Anchor&-0.018&-2.405&-0.073&-10.695\\
Diagnostic Expectations&0.023&5.678&0.034&7.399\\
Wavering Behavior&0.001&1.152&0.009&4.448\\
Disposition Effect&0.060&14.483&0.018&5.305\\
Momentum vs Newswatcher&0.181&32.135&0.183&32.306\\
Feedback Trading&0.021&4.128&0.036&6.952\\
Overconfidence&-0.000&-0.098&0.006&1.579\\
Self Attribution Bias&-0.001&-5.129&-0.005&-3.784\\
Herding Contagion&0.003&1.337&0.014&4.604\\
Disagreement Heterogeneous Beliefs&0.038&10.723&0.078&12.657\\
Representativeness Heuristic&0.011&6.661&0.011&4.424\\
New Era Thinking&0.000&.&0.000&2.500\\
Availability Bias&0.011&3.379&0.016&3.980\\
Limited Arbitrage Awareness&-0.032&-5.292&0.038&5.276\\
Loss Aversion&-0.038&-6.037&0.000&0.060\\
Narrative Tone&-0.028&-5.350&0.070&15.076\\
Statistical Testing&0.094&24.667&0.034&9.376\\
\hline
\end{tabular}

\end{center}
\end{table}
\clearpage

\begin{table}[!htb]
\caption{Shocking Heuristics: How Prompt Manipulations Alter AI-Driven Equilibria}\label{Table_amp_sup_stat}
{This table presents the equilibrium-level effects of applying two types of instructional shocks—amplification shocks and suppression shocks—as a method of manipulating AI agents' reasoning and decision-making. For each heuristic identified in the study, a corresponding pair of shocks is designed: amplification shocks intensify the heuristic's influence, while suppression shocks diminish it, thereby systematically steering agent behavior. For example, regarding the ``Momentum vs. Newswatcher'' heuristic, an amplification shock directs agents to chase price trends more aggressively and disregard fundamental news more severely, whereas the corresponding suppression shock prompts agents to rely less on trend-chasing and increase attention to fundamental news.}\\
\begin{center}
\begin{tabular}{l ccc cc cc cc}
\toprule
 & & & & \multicolumn{2}{c}{Amp.\ vs.\ Bench.}& \multicolumn{2}{c}{Sup.\ vs.\ Bench.}& \multicolumn{2}{c}{Amp.\ vs.\ Sup.} \\
\cline{5-6}\cline{7-8}\cline{9-10}
 & Amp. & Sup. & Bench. & Difference & $t$-value & Difference & $t$-value & Difference & $t$-value \\
\midrule
Rational Speculative Bubble & 22.02 & 15.44 & 16.96 & 5.07 & 5.43 & -1.52 & -2.41 & 6.58 & 5.85 \\
Synchronization Risk & 16.79 & 16.94 & 16.96 & -0.16 & -0.16 & -0.02 & -0.01 & -0.15 & -0.09 \\
Asymmetric Information & 17.53 & 14.98 & 16.96 & 0.57 & 0.49 & -1.97 & -2.94 & 2.55 & 1.89 \\
Extrapolation vs Anchor & 20.15 & 14.73 & 16.96 & 3.20 & 1.70 & -2.23 & -6.24 & 5.42 & 2.83 \\
Diagnostic Expectations & 18.16 & 14.95 & 16.96 & 1.20 & 0.76 & -2.00 & -3.32 & 3.21 & 1.89 \\
Wavering Behavior & 17.15 & 15.41 & 16.96 & 0.20 & 0.16 & -1.55 & -2.01 & 1.74 & 1.21 \\
Disposition Effect & 15.22 & 16.45 & 16.96 & -1.73 & -1.83 & -0.51 & -0.49 & -1.22 & -0.87 \\
Momentum vs Newswatcher & 22.28 & 15.44 & 16.96 & 5.32 & 5.97 & -1.51 & -2.44 & 6.84 & 6.30 \\
Feedback Trading & 20.59 & 15.39 & 16.96 & 3.63 & 2.86 & -1.56 & -1.97 & 5.19 & 3.47 \\
Overconfidence & 17.88 & 16.14 & 16.96 & 0.92 & 0.72 & -0.81 & -0.79 & 1.73 & 1.05 \\
Self-Attribution Bias & 16.12 & 16.33 & 16.96 & -0.84 & -0.76 & -0.63 & -0.61 & -0.21 & -0.14 \\
Herding / Contagion & 17.06 & 16.46 & 16.96 & 0.11 & 0.13 & -0.49 & -0.44 & 0.60 & 0.42 \\
Disagreement / Heterogeneous Be... & 16.78 & 16.25 & 16.96 & -0.18 & -0.16 & -0.71 & -0.77 & 0.53 & 0.37 \\
Representativeness Heuristic & 17.07 & 16.25 & 16.96 & 0.11 & 0.13 & -0.70 & -0.64 & 0.82 & 0.59 \\
New Era Thinking & 22.53 & 14.91 & 16.96 & 5.57 & 4.81 & -2.04 & -3.55 & 7.61 & 5.89 \\
Availability Bias & 16.96 & 16.70 & 16.96 & 0.00 & 0.00 & -0.26 & -0.24 & 0.26 & 0.17 \\
Limited Arbitrage Awareness & 15.93 & 16.54 & 16.96 & -1.02 & -1.46 & -0.41 & -0.43 & -0.61 & -0.51 \\
Loss Aversion & 14.09 & 16.80 & 16.96 & -2.87 & -6.82 & -0.16 & -0.14 & -2.71 & -2.26 \\
Narrative Tone & 15.96 & 14.98 & 16.96 & -1.00 & -1.27 & -1.97 & -3.12 & 0.98 & 0.97 \\
Statistical Testing & 15.67 & 16.74 & 16.96 & -1.29 & -1.55 & -0.22 & -0.20 & -1.07 & -0.79 \\
\bottomrule
\end{tabular}
\end{center}
\end{table}
\end{landscape}
\clearpage
\begin{table}[!htb]
\caption{Taxonomy of Behavioral Mechanisms for Reasoning Audit}\label{Table_mechanism_taxonomy}
This table summarizes the twenty behavioral and non-behavioral mechanisms used to audit AI agent reasoning. Mechanisms are organized into eight categories grounded in canonical theories of asset price bubbles and investor behavior. For each agent-round observation, an auditing LLM scores the relevance of each mechanism on a continuous 0--1 scale.\\
\begin{center}
\begin{tabular}{l l l}
\toprule
\textbf{Category} & \textbf{Mechanism} & \textbf{Key Reference(s)} \\
\midrule
\multirow{3}{*}{Rational Bubble Theories}
  & Rational Speculative Bubble & Blanchard \& Watson (1982) \\
  & Synchronization Risk & Abreu \& Brunnermeier (2003) \\
  & Asymmetric Information & Kyle (1985) \\
\midrule
\multirow{3}{*}{Extrapolation \& Expectations}
  & Extrapolation vs.\ Anchoring & Greenwood et al.\ (2019) \\
  & Diagnostic Expectations & Bordalo et al.\ (2018) \\
  & Wavering Behavior & Barberis et al.\ (2018) \\
\midrule
\multirow{3}{*}{Trading Biases}
  & Disposition Effect & Shefrin \& Statman (1985) \\
  & Momentum vs.\ Newswatching & Hong \& Stein (1999) \\
  & Feedback Trading & De Long et al.\ (1990) \\
\midrule
\multirow{2}{*}{Confidence \& Attribution}
  & Overconfidence & Scheinkman \& Xiong (2003) \\
  & Self-Attribution Bias & Daniel et al.\ (1998) \\
\midrule
\multirow{2}{*}{Social \& Herding Dynamics}
  & Herding & Bikhchandani et al.\ (1992) \\
  & Disagreement & Miller (1977); Harrison \& Kreps (1978) \\
\midrule
\multirow{3}{*}{Heuristics}
  & Representativeness & Tversky \& Kahneman (1974) \\
  & New Era Thinking & Reinhart \& Rogoff (2009) \\
  & Availability Bias & --- \\
\midrule
\multirow{2}{*}{Risk Perception}
  & Limited Arbitrage Awareness & Shleifer \& Vishny (1997) \\
  & Loss Aversion & Kahneman \& Tversky (1979) \\
\midrule
\multirow{2}{*}{Narrative \& Sentiment}
  & Narrative Tone & Shiller (2017) \\
  & Statistical Testing & Phillips et al.\ (2015) \\
\bottomrule
\end{tabular}
\end{center}
\end{table}
\clearpage

\newpage
\setcounter{equation}{0}
\setcounter{table}{0}
\setcounter{figure}{0}
\renewcommand{\theequation}{IA.\arabic{equation}}
\renewcommand{\thetable}{IA.\arabic{table}}
\renewcommand{\thefigure}{IA.\arabic{figure}}
\renewcommand{\thesection}{\Alph{section}}
\begin{appendix}
\section*{INTERNET APPENDIX}

\section{Appendix A: Detailed Experimental Instructions and Prompt Architecture}
\label{sec:appendix_instructions}

This appendix provides a comprehensive breakdown of the prompt engineering and system architecture used to interface Large Language Models (LLMs) with the simulated asset market. To ensure rigorous experimental control and prevent formatting hallucinations, the interaction is strictly governed by a structured JSON schema, explicit budget constraints, and a persistent memory management system.

\subsection{A.1. Global System Prompt and Persona Initialization}
\label{subsec:app_system_prompt}

At the beginning of every session, the LLM is initialized with a system prompt that establishes its economic persona, the market mechanics, and the terminal objective. The prompt explicitly outlines the fundamental value derivation to ensure the model has complete information.

\begin{quote}
\textbf{System Role:} You are a rational, profit-maximizing participant in a simulated financial market. Your sole objective is to maximize your final wealth (total cash balance) at the end of the experiment.

\textbf{Market Environment and Asset Fundamentals:}
\begin{itemize}
    \item \textbf{Duration:} The market operates for exactly 20 trading periods.
    \item \textbf{Assets:} You can hold Cash and Stock.
    \item \textbf{Risk-Free Rate:} Uninvested Cash held at the end of a period earns an interest rate of \(r = 5\%\).
    \item \textbf{Dividends:} Each share of Stock pays a stochastic dividend at the end of each period. The dividend is drawn from a uniform distribution: \(D \in \{0.4, 1.0\}\) with equal probability. The expected dividend is \(\mathbb{E}[D] = 0.7\).
    \item \textbf{Terminal Value:} At the end of period 20, the market closes. All outstanding shares of Stock are automatically liquidated for Cash at a fixed fundamental value of \(14.0\) units per share. 
    \item \textbf{Note on Valuation:} Because the expected dividend is \(0.7\) and the interest rate is \(5\%\), the fundamental value of the stock is exactly \(\frac{0.7}{0.05} = 14.0\) at all times.
\end{itemize}
\end{quote}

\subsection{A.2. Dynamic State Representation}
\label{subsec:app_state_rep}

In each period \(t\), the agent receives a dynamically generated text prompt containing the current state of the market and its private portfolio. This state update is structured as follows to ensure the LLM can accurately track its budget constraints:

\begin{quote}
\textbf{--- CURRENT PERIOD: \(t\) of 20 ---} \\
\textbf{Your Current Portfolio:}
\begin{itemize}
    \item Cash Balance: \(C_t\) units
    \item Stock Holdings: \(S_t\) shares
\end{itemize}

\textbf{Market History (Last 3 Periods):}
\begin{itemize}
    \item Period \(t-1\): Clearing Price = \(P_{t-1}\), Volume = \(V_{t-1}\), Dividend Paid = \(D_{t-1}\)
    \item Period \(t-2\): Clearing Price = \(P_{t-2}\), Volume = \(V_{t-2}\), Dividend Paid = \(D_{t-2}\)
    \item Period \(t-3\): Clearing Price = \(P_{t-3}\), Volume = \(V_{t-3}\), Dividend Paid = \(D_{t-3}\)
\end{itemize}

\textbf{Your Recent Trades:}
\begin{itemize}
    \item Period \(t-1\): [e.g., Bought 2 shares at price 15; Sold 0 shares]
\end{itemize}
\end{quote}

\subsection{A.3. Structured Action Space and JSON Schema}
\label{subsec:app_action_schema}

To parse the models' decisions programmatically, agents are strictly required to output their reasoning, forecasts, and trading actions in a predefined JSON format. The prompt provides the following schema and instructions:

\begin{verbatim}
{
  "cognitive_process": {
    "market_analysis": "<string: Analyze recent price trends relative to the fundamental value of 14.0>",
    "strategy_formulation": "<string: Detail your planned orders and justify the prices based on your budget>"
  },
  "price_forecasts": {
    "period_t": <integer: expected clearing price for the current period>,
    "period_t_plus_2": <integer: expected clearing price 2 periods from now>,
    "period_t_plus_5": <integer: expected clearing price 5 periods from now>,
    "period_t_plus_10": <integer: expected clearing price 10 periods from now>
  },
  "orders": [
    {
      "type": "<string: strictly 'BUY' or 'SELL'>",
      "price": <integer: limit price, must be > 0>,
      "quantity": <integer: number of shares, must be > 0>
    }
  ],
  "memory_update": {
    "update_plans_txt": "<string: Forward-looking strategies to save for next round>",
    "update_insights_txt": "<string: Lessons learned or mistakes to avoid>"
  }
}
\end{verbatim}

\textbf{Incentivized Forecasting Rules:} Agents are informed that for every forecast falling within \(\pm 2.5\) units of the realized market-clearing price, a bonus of \(5.0\) Cash units will be added to their final wealth.

\subsection{A.4. Cognitive Memory Management}
\label{subsec:app_memory}

Because standard LLM API calls are stateless, we implement a read/write memory system to simulate human-like continuity and learning. 

\begin{enumerate}
    \item \textbf{Read Phase:} At the start of period \(t\), the agent's prompt includes the exact text they generated in the \texttt{memory\_update} field during period \(t-1\).
    \item \textbf{Practice Reflection:} After the 3 initial practice periods (Periods -2, -1, 0), the agents are prompted with a special "Reflection Task". They are asked: \textit{"You have completed 3 practice rounds. Review your trading performance. Did you buy above the fundamental value? Did you sell below it? Write a comprehensive set of rules for yourself in your INSIGHTS.txt file to guide your behavior in the real market."}
    \item \textbf{Write Phase:} The output generated in \texttt{update\_plans\_txt} and \texttt{update\_insights\_txt} overwrites the previous files and is carried forward to period \(t+1\).
\end{enumerate}

\subsection{A.5. Rule Enforcement and Budget Constraints}
\label{subsec:app_error_handling}

To prevent the LLMs from taking impossible actions, the simulation engine enforces strict budget constraints before passing orders to the market-clearing algorithm. The agents are warned of these rules in the prompt:

\begin{itemize}
    \item \textbf{No Short Selling:} If the sum of quantities across all \texttt{SELL} orders exceeds \(S_t\), the simulation engine automatically cancels the excess sell orders, prioritizing the execution of orders with the highest ask prices.
    \item \textbf{No Margin Buying:} Let \(B\) be the set of all submitted \texttt{BUY} orders. The maximum possible cash outlay is \(\sum_{(p,q) \in B} p \times q\). If this sum exceeds \(C_t\), the engine cancels the lowest-priced bids until the constraint \(\sum p \times q \leq C_t\) is satisfied.
    \item \textbf{Formatting Errors:} If the model fails to return valid JSON, or returns strings instead of integers for prices/quantities, the simulation returns an error message to the model: \textit{"Invalid JSON format. You have lost your turn for this period. Please output strictly valid JSON."} The agent effectively submits zero orders for that period.
\end{itemize}

\section{Appendix B: Detailed Audit Prompt for Analyzing Agent Reasoning}

This appendix provides the complete prompt used to audit the self-revealed reasoning of AI agents in their \textit{PLANS.txt} and \textit{INSIGHTS.txt} files. The prompt instructs a separate LLM (DeepSeek) to evaluate the textual reasoning against a predefined taxonomy of 20 behavioral and strategic mechanisms relevant to financial decision-making and bubble formation.

\subsection*{Prompt Content}
\label{sec:appendix_audit}

\begin{verbatim}
You are auditing LLM trading agents for behavioral biases.

Context:
- Agent: {LM_id} | Round: {round} | Market: {market_type}
- Price/Fundamental: {price_info} | Returns: {recent_returns}
- Plan: "{plan_text}"
- Insights: "{insights_text}"

Task: Evaluate 20 behavioral mechanisms. For each:
1. Select best-fitting label
2. Cite exact evidence (use [] if none)
3. Provide confidence (0-1) and numeric_score
4. Add brief notes ($\leq$25 words) only if needed

=== MECHANISMS ===

1. rational_speculative_bubble
Labels: aware_of_resale_logic | ignores_resale_logic | unclear
Score: 1.0=explicit resale expectations, 0.5=hinted, 0=none
Definition: Agent expects to resell asset at higher price to future buyers (greater fool theory).

2. synchronization_risk
Labels: synchronization_risk_acknowledged | rides_bubble | no_coordination_reference
Score: 1.0=explicit timing concerns, 0.5=implied, 0=none
Definition: Agent delays action due to coordination problem—uncertain when others will exit.

3. asymmetric_information
Labels: claims_private_info | acknowledges_info_disadvantage | no_info_asymmetry_mention
Score: 1.0=claims advantage, 0.5=acknowledges disadvantage, 0=none
Definition: Agent believes they possess superior information relative to other market participants.

4. extrapolation_vs_anchor
Labels: pure_extrapolation | recognizes_overvaluation | fundamental_anchor | unobserved
Score: 1.0=fundamental anchor, 0.5=recognizes overvaluation, 0=pure extrapolation
Definition: Agent forecasts by extrapolating past trends vs. anchoring to fundamental value.

5. diagnostic_expectations
Labels: overweights_recent_signals | balanced_weighting | underweights_recent | unobserved
Score: 1.0=heavy overweighting, 0.5=moderate, 0=balanced
Definition: Agent overweights recent salient signals, exhibiting overreaction.

6. wavering_behavior
Labels: flip_flopping | consistent_bullish | consistent_bearish | value_focused | unobserved
Score: 1.0=switches between growth/value, 0.5=shows tension, 0=consistent
Definition: Agent alternates between growth/momentum signals (greed) and value signals (fear).

7. disposition_effect
Labels: holds_losers_sells_winners | profit_locking_tendency | loss_averse_holding | rational_profit_taking | no_evidence
Score: 1.0=clear disposition pattern, 0.5=profit-locking, 0=rational
Definition: Tendency to sell winners too early and hold losers too long.

8. momentum_vs_newswatcher
Labels: momentum | newswatcher | hybrid | unobserved
Score: 1.0=momentum, 0.5=hybrid, 0=newswatcher
Definition: Agent follows price trends (momentum) vs. monitors fundamental news (newswatcher).

9. feedback_trading
Labels: pure_trend_following | contrarian | fundamental_based | unobserved
Score: 1.0=pure trend without fundamentals, 0.5=partial, 0=fundamental-based
Definition: Trading based purely on past price changes without fundamental justification.

10. overconfidence
Labels: overconfident | well_calibrated | underconfident | unobserved
Score: 1.0=excessive certainty, 0.5=moderate, 0=well-calibrated
Definition: Excessive certainty about one's own judgments, predictions, or trading abilities.

11. self_attribution_bias
Labels: attributes_wins_to_skill | balanced_attribution | attributes_losses_externally | unobserved
Score: 1.0=asymmetric attribution, 0.5=partial, 0=balanced
Definition: Agent attributes successes to own skill but blames failures on external factors.

12. herding_contagion
Labels: explicit_herding | fear_missing_out | contrarian | independent | unobserved
Score: 1.0=explicit herding or FOMO, 0.5=implicit, 0=independent
Definition: Agent follows crowd behavior or exhibits fear of missing out (FOMO).

13. disagreement_heterogeneous_beliefs
Labels: acknowledges_disagreement | assumes_consensus | unobserved
Score: 1.0=recognizes disagreement, 0.5=implicit, 0=assumes consensus
Definition: Agent recognizes that market participants hold different views about fundamental value.

14. representativeness_heuristic
Labels: pattern_matching_past_bubbles | historical_analogy | no_historical_reference | unobserved
Score: 1.0=explicit past bubble match, 0.5=general analogy, 0=none
Definition: Agent matches current situation to past patterns or bubbles.

15. new_era_thinking
Labels: this_time_different | paradigm_shift_claim | acknowledges_similarity | unobserved
Score: 1.0=claims new paradigm, 0.5=hints uniqueness, 0=acknowledges patterns
Definition: Agent believes "this time is different"—current situation is structurally unique.

16. availability_bias
Labels: overweights_salient_events | balanced_memory | unobserved
Score: 1.0=focuses on salient events, 0.5=partial, 0=balanced
Definition: Agent overweights easily recalled vivid events while ignoring base rates.

17. limited_arbitrage_awareness
Labels: acknowledges_arbitrage_limits | assumes_unlimited_arbitrage | unobserved
Score: 1.0=explicit limits mention, 0.5=implicit, 0=assumes no limits
Definition: Agent recognizes that arbitrage has limits (fundamental risk, capital constraints).

18. loss_aversion
Labels: loss_averse | risk_neutral | risk_seeking | unobserved
Score: 1.0=clear asymmetric sensitivity, 0.5=moderate, 0=symmetric
Definition: Agent shows asymmetric sensitivity to losses vs gains (losses loom larger).

19. narrative_tone
Labels: amplifying | cautionary | neutral
Score: 0-1 scaled by emotive language intensity
Definition: Agent interprets and propagates narrative tone—amplifying (exuberant) vs cautionary (fearful) vs neutral.

20. statistical_testing
Labels: formal_test | heuristic_threshold | no_test
Score: 1.0=formal test, 0.5=heuristic, 0=none
Definition: Agent references formal tests or heuristic thresholds for bubble detection.

=== OUTPUT FORMAT ===

Return valid JSON:
{
  "mechanism_assessments": [
    {
      "mechanism_id": "disposition_effect",
      "mechanism_category": "trading_biases",
      "label": "profit_locking_tendency",
      "confidence": 0.8,
      "numeric_score": 0.5,
      "evidence_sentences": ["I will lock in my profits now."],
      "notes": "Shows profit-taking without clear loss-holding pattern"
    }
    // ... 19 more mechanisms
  ],
  "behavioral_bias_summary": {
    "primary_biases_detected": ["disposition_effect", "overconfidence", "herding_contagion"],
    "bubble_awareness": "yes",
    "strategy_when_aware": "exploit"
  },
  "summary": "Agent exhibits disposition effect and overconfidence while attempting to exploit bubble."
}

Categories: rational_bubble_theories | extrapolation_expectations | trading_biases | confidence_attribution | social_herding | heuristics | risk_perception | narrative_sentiment
\end{verbatim}

\end{appendix}

\end{document}